\documentclass[12pt]{iopart}
\usepackage{iopams}
\newcommand{\eqref}[1]{(\ref{#1})}
\usepackage{amssymb}
\usepackage{hyperref}
\usepackage{tabularx}
\usepackage[dvipsnames]{xcolor}
\usepackage{soul}
\usepackage{siunitx} %
\usepackage{selinput}%
\usepackage{todonotes}
\usepackage[capitalise]{cleveref} %
\usepackage{upgreek} %

\usepackage{graphicx}
\usepackage{blindtext}
\def\newblock{\ } %
\usepackage{tabulary}
\usepackage{rotating}
\usepackage{ragged2e} %
\usepackage{afterpage}

\usepackage{multirow}
\usepackage{makecell}

\renewcommand{\thesection}{\arabic{section}}
\renewcommand{\thesubsection}{\thesection.\arabic{subsection}}
\renewcommand{\thesubsubsection}{\thesubsection.\arabic{subsubsection}}
\makeatletter
\newcommand*{\biggg}[1]{{\hbox{$\left#1\vbox to20.5\p@{}\right.\n@space$}}}
\newcommand*{\Biggg}[1]{{\hbox{$\left#1\vbox to23.5\p@{}\right.\n@space$}}}
\newcommand*{\Bigggg}[1]{{\hbox{$\left#1\vbox to26.5\p@{}\right.\n@space$}}}
\renewcommand{\p@subsection}{}
\renewcommand{\p@subsubsection}{}
\def\p@paragraph{\thesubsection.}
\makeatother

\Crefname{section}{Sec.}{Secs.}

\begin{document}
\title[Non-geometric tilt-to-length coupling in precision interferometry]{Non-geometric tilt-to-length coupling in precision interferometry: mechanisms and analytical descriptions}

\author{Marie-Sophie Hartig, S\"onke Schuster, Gerhard Heinzel, and Gudrun Wanner}

\address{Max Planck Institute for Gravitational Physics (Albert Einstein Institute) and 
Institute for Gravitational Physics of the Leibniz Universit\"at Hannover, Callinstrasse 38, 30165 Hannover, Germany}

\ead{\mailto{marie-sophie.hartig@aei.mpg.de}, \mailto{gudrun.wanner@aei.mpg.de}}
\vspace{10pt}
\begin{indented}
\item[]November 2021
\end{indented}

\begin{abstract}
This paper is the second in a set of two investigating tilt-to-length (TTL) coupling.
TTL describes the cross-coupling of angular or translational jitter into an interferometric phase signal and is an important noise source in precision interferometers, including space gravitational wave detectors like LISA. 
We discussed in \cite{G21} the TTL coupling effects originating from optical path length changes, i.e. geometric TTL coupling.
Within this work, we focus on the wavefront and detector geometry dependent TTL coupling, called non-geometric TTL coupling, in the case of two interfering fundamental Gaussian beams.
We characterise the coupling originating from the properties of the interfering beams, i.e.\ their absolute and relative angle at the detector, their relative offset and the individual beam parameters. 
Furthermore, we discuss the dependency of the TTL coupling on the geometry of the detecting photodiode. 
Wherever possible, we provide analytical expressions for the expected TTL coupling effects.
We investigate the non-geometric coupling effects originating from beam walk due to the angular or translational jitter of a mirror or a receiving system.
These effects are directly compared with the corresponding detected optical path length changes in \cite{G21}. 
Both together provide the total interferometric readout.
We discuss in which cases the geometric and non-geometric TTL effects cancel one-another. 
Additionally, we list linear TTL contributions that can be used to counteract other TTL effects.
Altogether, our results provide key knowledge to minimise the total TTL coupling noise in experiments by design or realignment.
\end{abstract}
\noindent{\it Keywords\/}: tilt-to-length coupling, optical cross-talk, wavefront properties, interferometric noise sources, laser interferometry, space interferometry, LISA

\section{Introduction}
Tilt-to-Length (TTL) coupling is a common type of noise in precision laser interferometers. It describes the unwanted coupling of angular or translational jitter into the phase readout. 
The jittering object can, for example, be a reflective component or a receiving system, i.e.\ an optical bench or a satellite that is receiving the beam of interest.

In general, TTL coupling originates, on the one hand, from the fact that the optical distance along the beam axis is lengthened or shortened by the jitter. 
This results in changes of the optical path length difference (OPD) of the beam axis and has been discussed in detail in \cite{G21}. 
On the other hand, the jitter changes the interference pattern. These changes depend on the wavefront properties and originate from alterations in the beam alignment with respect to the detector surface and with respect to each other. Additionally, beam displacements on the detector surface can result in beam clipping by the boundaries of the detector surface. All these effects contribute to the final interferometric output signal. 
After all, an interferometer does not directly sense an OPD but rather a phase difference between interfering beams on a detector surface. This is typically converted to a length readout signal, the longitudinal path length sensing (LPS) signal, by dividing the phase by the wavenumber $k$ \cite{Chwalla2020,wanner2012methods}.
We categorise these LPS changes due to the interfering beams' wavefront properties and the detector geometry as non-geometric effects.

Although it is often sufficient to focus on the geometric TTL effects, %
there are cases where the non-geometric TTL effects become equal or even dominant noise contributors. 
One example is given by setups in which the centre of rotation is located in the beam's point of incidence on the detector, which results in the suppression of the geometric coupling effects.
This is implemented in the LISA mission \cite{LISAMission,Jennrich2009} by imaging systems 
\cite{Chwalla2016,Chwalla2020} reducing the geometric jitter coupling of the receiving spacecraft. 
Another important example of dominant non-geometric coupling is the TTL coupling of the jitter of a transmitting spacecraft in the LISA mission into the long arm interferometer readout, computed, for instance, in \cite{sasso2018misalignment,sasso2018far-field,sasso2019}, which we do not consider in this work. 
More general examples that are independent of a specific mission have been introduced in \cite{G21,wanner2012methods,Schuster2015}, and are further discussed within this paper. 

TTL coupling has been investigated for different missions in various publications.
It is discussed for the LISA long arm interferometer in \cite{Chwalla2020,sasso2018misalignment,sasso2018far-field,sasso2019}, for the LISA test mass interferometer in \cite{Troebs2018}, and for the GRACE-FO mission in \cite{Henry2020}, and for the LISA Pathfinder mission in \cite{armano2016sub,Wanner2017,armano2018cali,armano2019lpf}.
Here, we provide a very general overview of non-geometric TTL effects applicable to precision interferometers, including space interferometers. 
The work presented here is fundamental and not limited to a dedicated mission or project.

The scope of this work is to present analytical descriptions for a series of non-geometric TTL coupling effects and to characterise these as first- or second-order coupling effects. 
Therefore, our manuscript is in terms of content divided into two parts. 
We start with defining the non-geometric TTL coupling effects analytically for fundamental Gaussian beams and local interferometric effects (Sec.~\ref{sec: wavefront related ttl}).
In the following sections, we then evaluate the analytical descriptions of the non-geometric TTL coupling depending on the wavefront properties (Sec.~\ref{sec: wavefront geometry}) and detector geometry (Sec.~\ref{sec: photodiode geometry}).
Thereby, we assume one of the beams to jitter either angularly or translationally while the other one remains static. 

In the second half, the results are interpreted for two common jittering objects in Sec.~\ref{sec:BeamWalkMechanisms}: a receiving system and a reflective component.
The presented equations are complementary to the geometric coupling presented in \cite{G21}.
Note that TTL coupling from a jittering transmitter is not specifically discussed here.
If the transmitter is close to the receiver, e.g., in a laboratory setup, this case is equivalent to the case of a jittering mirror with the point of reflection being the point of rotation.
If, otherwise, the transmitter is far from the receiver, the beam diameter would have increased to a width that the receiving aperture cannot fully capture. 
Hence it would be clipped and the Gaussian beam assumption, which forms the basis of this paper, would be no longer applicable. 
In that case, other simulation methods, such as those discussed in  \cite{Zhao2022,sasso2018misalignment,SchusterPhD} are needed to account for diffraction effects and also wavefront errors.

We summarise all non-geometric effects in Sec.~\ref{sec:summary} and list them for a typical special case.
In Sec.~\ref{sec: overview}, we extend this summary to the full LPS signal. There, we combine the results from \cite{G21} and this work to discuss the total signal.
Finally, we give a conclusion in Sec.~\ref{sec:conclusion}.

\section{The definition of non-geometric TTL coupling}
\label{sec: wavefront related ttl}
For the introduction into the topic, we explain in this section how non-geometric TTL coupling is defined and how it relates to geometric coupling contributions (Sec.~\ref{sec: non geom analytic derivation}).
Furthermore, we present how it can be derived analytically and in numerical simulations (Sec.~\ref{sec:General_Settings_for_LPS_ng}).  

\subsection{Introduction: LPS and OPD}
\label{sec: non geom analytic derivation}
Throughout this paper, we assume  the case of interfering Gaussian beams, which can be described according to \cite{wanner2012methods} by
\begin{eqnarray}
	E(r_b,z_b,t)\propto& \frac{1}{w(z_b)}\exp\left(\frac{-r_b^2}{w^2(z_b)}\right)\cdot \nonumber\\
	& \exp\left(i \Omega t - i\left[ \frac{k r_b^2}{2 R(z_b)} - \zeta (z_b) + k \,\text{OPL}_b\right]\right)\;,
\label{eq:electric_field}
\end{eqnarray}
where all variable definitions are listed in Tab.~\ref{tab:Variables}. %
The lower index $b$ of Eq.~\eqref{eq:electric_field} stands for `beam' and is, therefore, substituted in Tab.~\ref{tab:Variables} by either $m$ or $r$ referring to a measurement or reference beam respectively. The measurement beam is the beam of interest that is subject to the jitter.
This beam interferes with a second beam, the reference beam, on the detector. This reference beam is assumed to be static with respect to the detector surface. 
\fulltable{List of physical parameters.}
  \label{tab:Variables}
  \centering
\begin{tabularx}{\textwidth}{l  X  p{3.3cm}}
	\br
	parameter		& description & characterising eq. \\
	\mr
	$k$			& wave number common for both beams	& $k= 2 \pi/\lambda$ \\
	$\lambda$	& wavelength &\\
	$\Omega_{m,r}$ & angular frequency of meas.\ / ref.\ beam & $\Omega=c k = 2 \pi f $\\
	$\Delta \Omega$ & angular heterodyne frequency & $\Delta \Omega=| \Omega_m-\Omega_r |$ \\
	$z_{R,m,r}$	& Rayleigh range of meas.\ / ref.\ beam & $z_R= \pi \,w_0^2/\lambda$\\ 
	$w_{0,m,r}$	& waist size of meas.\ / ref.\ beam &$w_0= \sqrt{z_R\, \lambda/\pi}$ \\
	$w_{m,r}$		& laser spot size on detector & $ w=w_0 \sqrt{1+ (z/z_R)^2}$\\
	$R_{m,r}$ 	& radius of curvature of meas.\ / ref.\ beam & $R=z\, (1+(z_R/z)^2)$\\
	$\zeta_{m,r}$	& Gouy phase of meas.\ / ref.\ beam & $\zeta = \arctan(z/z_R)$\\
	$P_{m,r}$		& power of meas.\ / ref.\ beam &\\
	$z_{m,r}$		& distance from waist in direction of propagation for meas. / ref. beam &\\
	$s_{m,r}$		& propagation distance of meas.\ / ref.\ beam &\\
	$r_{m,r}$		& cylindrical coordinate  of meas.\ / ref.\ beam& $r=\sqrt{x^2+y^2}$\\
	OPL$_{m,r}$		& optical path length of the meas.\ / ref.\ beam axis  & \\
	\br
\end{tabularx}
\endfulltable

For two interfering beams, the detected power can be derived via the integral of the squared absolute sum of their electric fields over the photodiode surface $S$:
\begin{eqnarray}
P &\propto  \int dS\, \Vert E_m + E_r\Vert^2 \\
&=  \int dS\, \left(\Vert E_m\Vert^2 + \Vert E_r\Vert^2 + E_m E_r^* + E_m^* E_r\right) \, .
\end{eqnarray}
The first two summands describe the detected power of the individual beams.
The third and the fourth summand, evaluated for $t=0$, are often referred to as complex amplitudes $a$ of the beat note.
The argument of this complex amplitude is then the interferometric phase $\phi$ \cite{wanner2012methods}, e.g.
\begin{eqnarray}
	\phi = \arg(a) = \arg\left( \int dS \left.E_m E_r^*\right\vert_{t=0}\right)\,, 
	\label{eq:ifo_phase}
\end{eqnarray}
sensed by the corresponding interferometric readout system. 
This description is equally valid for homodyne and heterodyne interferometers.

The longitudinal path length sensing (LPS) signal is this phase converted to a length by a division by the wavenumber $k$:
\begin{equation}
	\text{LPS} = \frac{1}{k}\, \phi \, . \label{eq:LPS}
\end{equation}
This derivation is valid if the entire interference pattern is detected with homogenous sensitivity, so for single element photodiodes (SEPDs), which are sufficiently large to detect the complete incident wavefronts. We will extend this for quadrant photodiodes (QPDs) in \cref{sec: photodiode geometry}.

The LPS represents the actual displacement measurement of interferometers. 
While its geometric coupling contributor, i.e.\ the optical path length difference (OPD) of the two beam axes,
\begin{eqnarray}
  \text{OPD} = \text{OPD}_m - \text{OPD}_r \,,
\end{eqnarray}
can be derived from simple geometry \cite{G21},
the LPS is more complicated to compute both numerically and analytically. 
Because each OPL is defined along the corresponding beam axis, the OPD is (approximately) constant in the surface integral of Eq.~\eqref{eq:ifo_phase}. 
It can therefore be drawn out of this integral:
\begin{eqnarray}
	\phi&= \arg\bigg[\exp(- i k (\text{OPL}_m-\text{OPL}_r))\cdot \nonumber\\
	& \qquad\left( \int dS (E_m  E_r^*) \vert_{\text{OPL}_{m,r}=t=0} \right)\bigg] \label{eq:separate_macro-phase}\\
	&= \arg\bigg[ \exp(i k \,\text{OPD}) \left( \int dS (E_m  E_r^*) \vert_{\text{OPL}_{m,r}=t=0} \right) \bigg] \\
	&= k\, \text{OPD} + \arg\bigg[ \left( \int dS (E_m  E_r^*) \vert_{\text{OPL}_{m,r}=t=0} \right) \bigg].
	\label{eq:separate_macro-phase2}
\end{eqnarray}
Strictly speaking, this derivation only holds for beams of normal incidence. In the case of tilted beams, an additional microscopic phase needs to be considered when the OPD is taken out of the integral \cite{wanner2012methods}.
Yet, the OPD can be separated in either case from the non-geometric phase (second summand in Eq.~\eqref{eq:separate_macro-phase2}).
Thus, the LPS can be split into the OPD and a non-geometric contribution LPS$_\text{ng}$:
\begin{equation}
	\text{LPS} = \text{OPD} + \text{LPS}_\text{ng} \, .
	\label{eq:LPS+OPD+LPSng}
\end{equation} 
The non-geometric part LPS$_\text{ng}$ contains then all contributions related to the wavefront properties (i.e.\ the involved radii of curvature $R_b$, Gouy phases $\zeta_b$ and spot sizes $w_b$) and the beam alignment at the detector described by the angles ($\varphi_b,\eta_b$) and the point ($x_{ib},y_{ib}$) of incidence at the detector.
Additionally, it contains clipping effects if the detector surface $S$ is not large enough to receive the full extent of the impinging wavefronts. 
A more detailed description of the numerical implementation of these equations and possibly needed coordinate transformation is given in \cite{wanner2012methods}. The LPS signal measured in a laboratory experiment includes additional effects that are not included in the presented equations, such as the non-uniformity of the photodiode's responsivity.

\subsection{Derivation of LPS$_\text{ng}$ in analytical and in numerical simulations} 
\label{sec:General_Settings_for_LPS_ng}
In numerical simulations that allow both, a computation of the LPS as well as the OPD (this is, for instance, the case in IfoCAD \cite{wanner2012methods,Kochkina2013}), one can naturally derive the LPS$_\text{ng}$ simply from the difference
\begin{equation}
	\text{LPS}_\text{ng} = \text{LPS} - \text{OPD} \, .
	\label{eq:LPSng+LPSng-OPD}
\end{equation} 
In analytical derivations, this process would be significantly more complex than necessary. 
In the analytic and numeric approach, the OPD can be derived as described in \cite{G21} via the difference between the OPLs of the measurement beam in the tilted and the nominal case. 
However, the computation of the non-geometric TTL contributions can be simplified by adapting the simulation to directly exclude OPL changes and making them zero by design. 
In that case, the LPS signal in Eq.~\eqref{eq:LPSng+LPSng-OPD} is fully non-geometric and can be derived analytically using the procedure described in \cite{Wanner2014} summarised in the following:

\begin{figure}
	\centering
	\includegraphics[width=0.45\textwidth]{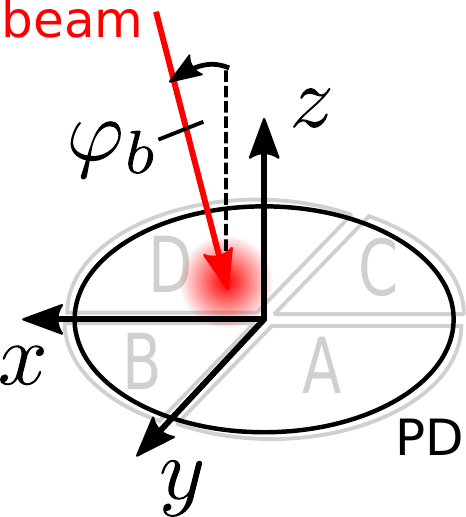}
	\caption{Coordinate system of a single element (black circle) or quadrant (grey quarter circles) photodiode, respectively. The four quadrants of the quadrant photodiode are labelled A, B, C and D.
	A beam rotation in the $xz$-plane is described by the angle $\varphi_b$. We do not show here rotations in the $yz$-plane ($\eta_b$).
	\label{fig:PD_coordsyst}
}
\end{figure}

To account for the tilt angle of each of the beams $b$ and the rotation axis defining a rotation matrix $M_{\text{rot},b}$, as well as the location of the centre of rotation $\textbf{p}_{\text{pivot},b}$,
we first perform a coordinate transformation in Eq.~\eqref{eq:electric_field}. 
The new coordinates are defined with respect to the photodiode centre, i.e.\ $z=0$, and in the detector coordinate system (Fig.~\ref{fig:PD_coordsyst}).
\begin{eqnarray}
\left(\hskip -\arraycolsep \begin{array}{*{20}{c}} 
x_b \\ y_b \\ \Delta\text{OPL}_b  
\end{array} \hskip -\arraycolsep\right)
= M_{\text{rot},b} \left[
\left(\hskip -\arraycolsep \begin{array}{*{20}{c}} 
x \\ y \\ 0
\end{array} \hskip -\arraycolsep\right)
- \textbf{p}_{\text{pivot},b} \right] 
+ \left[ \textbf{p}_{\text{pivot},b} -
\left(\hskip -\arraycolsep \begin{array}{*{20}{c}} 
x_{i,b} \\ y_{i,b} \\ 0
\end{array} \hskip -\arraycolsep\right)
\right]
\end{eqnarray}
The beams are then superimposed on the photodiode, and the overlap integral in Eq.~\eqref{eq:ifo_phase} and the resulting LPS signals are analytically evaluated. 
In the analytical evaluation of the overlap integral, it is assumed that the beam parameters do not vary for different detector points:
Due to the typically different shapes of the wavefront and the detector surface, the outer parts of the wavefront would hit the detector earlier or later, yielding slightly different parameters. However, the differences are small and scaled by a decreasing intensity. Hence, we only consider their values in their point of incidence.
This simplifying assumption is not necessary for numerical computations using IfoCAD. 
Also, in our comparisons between the analytical and the simulation results, we find no differences exceeding the accuracy achievable with IfoCAD ($\sim$10$^{-15}$\,m).

For the computation of the non-geometric signal contribution only, we substitute the pivot point $\textbf{p}_{\text{pivot},b}$ by the measurement beam's point of incidence $(x_\text{im}, y_\text{im},0)$ on the photodiode surface. 
By this replacement, all OPD contributions become zero and, therefore, $\text{LPS} =\text{LPS}_\text{ng}$.

In general, we reduce the complexity of the analytic equations by neglecting the jitter dependency of beam parameters as a higher-order contribution.

\subsection{Taylor expansion of the equations}
\label{sec:readme}
The evaluation of the non-geometric coupling equations, as described above, yields very complex terms, which make it hard to deduce the characteristics of the TTL coupling effects.
Therefore, we will show only the second-order series expansions in the following section.
The equations are Taylor-expanded in all small parameters ($\alpha\ll 1\,\text{rad}$ and $r\ll 1\,\text{m}$), which are the beam alignment angles ($\varphi_m,\,\eta_m,\,\varphi_r,\,\eta_r$), detector alignments ($\varphi_\text{PD},\,\eta_\text{PD}$) and shifts due to the translational jitter ($x_m,\,y_m$).

\section{Wavefront related TTL coupling effects: case of ideal detectors}
\label{sec: wavefront geometry}
In this section, we investigate how the wavefront properties, i.e.\ the beam parameters, affect the phase signal and show how this wavefront-dependent TTL coupling depends on the point of incidence at the detector.
To differentiate between the wavefront-related TTL effects 
and the detector-related TTL effects (Sec.~\ref{sec: photodiode geometry}), we assume in our analytic investigations here that the detectors are ideal infinitely large single element photodiodes (SEPD). 

We start with analysing the TTL coupling in the most simplified setup of two perfectly identical fundamental Gaussian beams with the measurement beam rotating around its point of incidence on the detector. 
We distinguish the case where the beams overlap perfectly (Sec.~\ref{sec: non_geom equal}) and the case of a shifted measurement beam Sec.~\ref{sec: non_geom offset equal beam}. 
Since identical Gaussian beams are idealisations impossible to achieve in reality, we further discuss TTL effects induced by beams having unequal beam parameters in Sec.~\ref{sec: non_geom beam parameter}.
Since any other distortion of the wavefront or intensity shape will alter the balance between wavefronts, it also generates cross-coupling, see Sec.~\ref{sec: higher oder modes}.

\subsection{Identical Gaussian beams and rotation around the common point of incidence}
\label{sec: non_geom equal}
\begin{tabular}{|ll|}
\br
jitter type: & $\varphi_m$  \\
static parameters: & $\varphi_r$, $\varphi_\text{PD}$, $z_{Rm}=z_{Rr}$, $z_m=z_r$, $x_{im}=x_{ir}$ \\
detector: &  SEPD \\
\br
\end{tabular}

We assume the most simple case of two identical fundamental Gaussian beams with identical points of incidence and a rotation of the measurement beam around their common point of incidence.
For a rotation angle of $\varphi_m=0$ both beams and beam paths are identical and they perfectly overlap each other. 
If rotated around the beams' shared incidence point on the detector, the measurement beam's geometric path length does not change, giving here a pure non-geometric path length readout.
In the following, we will derive the TTL coupling in this particular case first qualitatively via a graphic and then analytically. 

\begin{figure}
	\centering
	\includegraphics[width=0.9 \columnwidth]{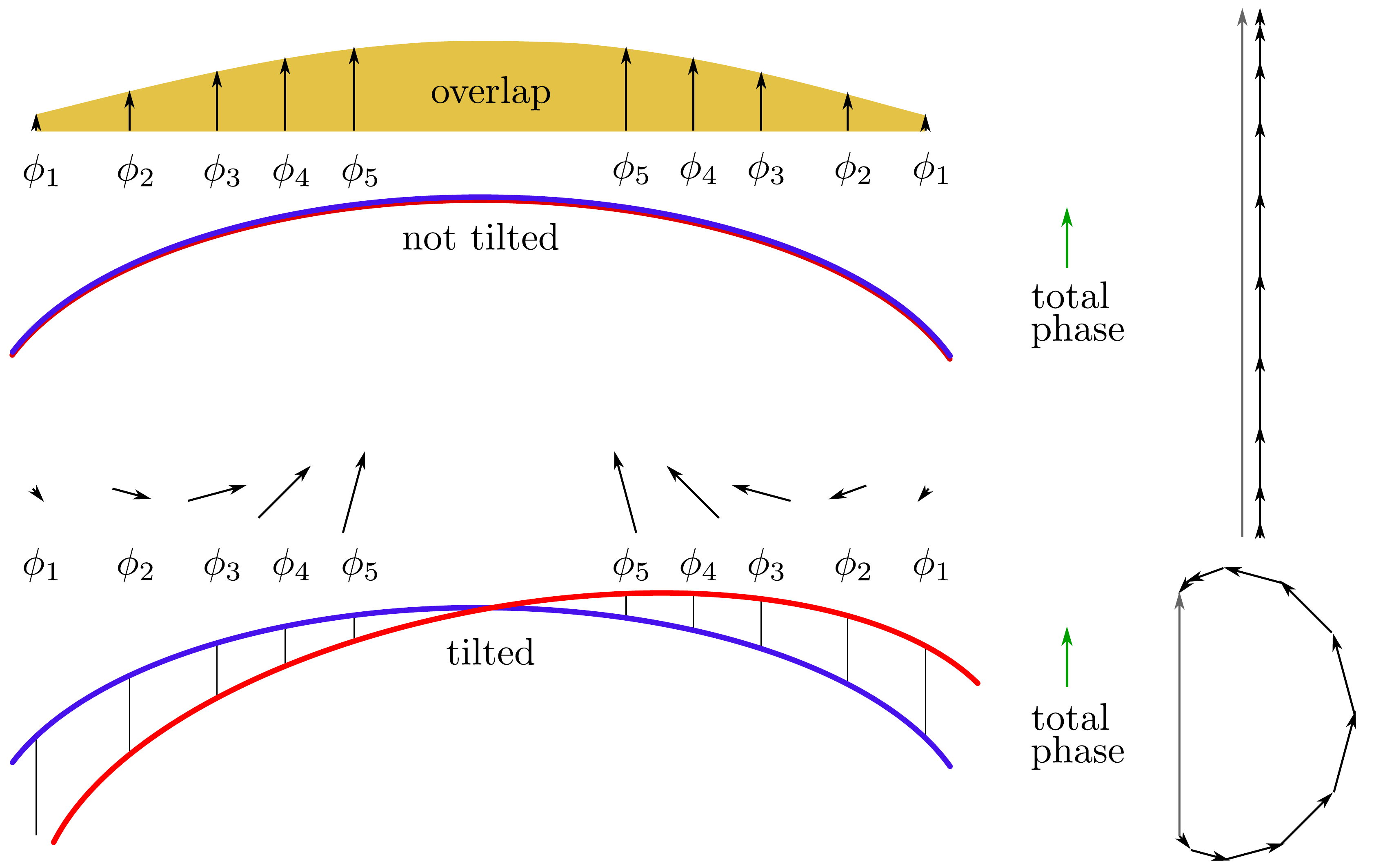}
	\caption{Simplified illustration for the vanishing TTL in case of identical Gaussian beams rotating around their joint point of incidence on the photodiode. Shown are the superposition of two wavefronts in the nominal untilted case (upper sketch with superimposing blue and red wavefronts) and the case when the measurement wavefront is tilted around the centre of the reference wavefront (lower sketch). The local phase differences between the wavefronts are indicated by $\phi_i$ and are illustrated by the directions of the black arrows (vector representations of the local complex amplitudes).
	The phase information at different positions on the detector is weighted with the amplitude of the total field (``overlap'') resulting from the superposition of the two interfering beams.
	The phase sum $\phi_\text{sum}$ represents the integrated phase over the entire detector. This $\phi_\text{sum}$ is indicated by the direction of the green arrows on the right-hand side. Due to the given symmetry, $\phi_\text{sum}$ is unaffected by the rotation, which shows that there is no non-geometric TTL, i.e.\ LPS$_\text{ng}=0$.}
	\label{fig: beam walk}
\end{figure}

The graphical derivation of TTL coupling is based on an approximation of Eq.~\eqref{eq:ifo_phase}. 
Therefore, we first divide the surface integral in Eq.~\eqref{eq:ifo_phase} into $n$ segments $S_i$. 
The complex amplitude $a$ then equals the sum over the complex amplitudes $a_i$ of the segments $S_i$
\begin{eqnarray}
	a =& \sum_{i=1}^n a_i \\
	a_i \propto& \int dS_i (E_m E_r^*) \vert_{t=0} \label{eq:Def_ComplexAmp_i}\;.
\end{eqnarray}
The expected interferometric phase $\phi$ is then given by the argument of the sum of all complex amplitudes:
\begin{equation}
	\phi = \arg(a) = \arg\left(\sum_{i=1}^n a_i\right) \,.
	\label{eq:phi_graphic}
\end{equation}

In a graphical approach, we examine a small number of complex amplitudes $a_i$ visually from the differences of the interfering wavefronts. 
This is illustrated in Fig.~\ref{fig: beam walk} showing the wavefronts of two interfering beams in the nominal non-tilted case and for an arbitrarily chosen tilt angle $\varphi_m$. 
Here, the vector representations of the complex valued amplitudes $a_i$ are given by the black arrows. 
Their directions define the local phase differences $\phi_i$.
In the figure, these are graphically estimated from the difference (averaged difference within a certain segment) of the phase profiles, i.e.\ the blue and red wavefronts.
Further, the lengths of the arrows are defined by the Gaussian amplitude profile of the interference pattern, qualitatively described by the yellow area.
In accordance with Eq.~\eqref{eq:phi_graphic}, the full complex amplitude equals the sum of the black arrows.
The interferometric phase $\phi$ then equals the angle of this vector sum. 

Next, we evaluate the total phases for the case shown in Fig.~\ref{fig: beam walk}:
While the local phase differences are all zero in the non-tilted case, a tilt of the measurement beam (red wavefront) changes the directions but not the lengths of the complex amplitude vectors. Due to the given symmetry, the arrows on the right- and left-hand sides are antisymmetric, which results again in a total phase of zero, i.e.\ the same value as in the non-tilted case, showing that no TTL coupling will occur. 

We confirm this graphical derivation by deriving the LPS signal analytically using the methods described in \cite{wanner2012methods, Wanner2014}.
Therefore, we evaluate Eq.~\eqref{eq:ifo_phase} for the electrical fields of identical Gaussian beams, i.e.\ beams with the same Rayleigh ranges $z_{Rm}=z_{Rr}$, distances from waist $z_{m}=z_{r}$ and incidence points $x_{im}=x_{ir}$ at the detector.
For small beam tilts $\varphi_m$ we find %
\begin{eqnarray}
\text{LPS}^\text{SEPD,2D}_\text{ng} \approx& - \frac{z_m}{4k\,z_{Rm}} \left( \varphi_m^2-2\varphi_m\varphi_\text{PD}\right) \,
\label{eq: non-geom equal}
\end{eqnarray}
which corresponds to results from \cite{Schuster2015} but additionally describes the contribution of a small tilt $\varphi_\text{PD}$ of the detector surface.

The term we find in the analytical derivation is assumed to originate from the simplifying assumptions made in the computation of the LPS signal (see Sec.~\ref{sec:General_Settings_for_LPS_ng}). 
For a small photodiode tilt $\vert\varphi_\text{PD}\vert \lesssim 200\,\upmu\mathrm{rad}$ and common parameters, e.g.\ $\vert\varphi_m\vert \lesssim 200\,\upmu\mathrm{rad}$, $\lambda=1064\,\mathrm{nm}$, $z_m\sim 1\,\mathrm{m}$, $z_{Rm}\sim 1\,\mathrm{m}$, it is smaller than $10^{-14}$\,m, and therefore few orders of magnitude smaller than usually needed in space-based interferometers.
Also, the computation accuracy of our comparison tool IfoCAD, while highly accurate in the case of beam tracing and geometric effects, is in its current version confined by the numerical limits of the integration algorithms ($\sim$10$^{-15}$m for typical simulations).
Thus, despite no coupling being visible in our simulations, the existence of residual coupling terms in the case of two identical Gaussian beams with one rotating around the shared point of incidence cannot fully be ruled out.
In either case, we can conclude that the TTL coupling for this scenario is negligible. 

Note that we initially assumed that the reference beam is not tilted.
However, also for a static misalignment by a small angle $\varphi_r$, Eq.~\eqref{eq: non-geom equal} would still hold:
The angular alignment of the reference beam does not couple with the measurement beam jitter.

\subsection{Identical Gaussian beams and rotation around a laterally shifted point of incidence}
\label{sec: non_geom offset equal beam}
\begin{tabular}{|ll|}
\br
jitter type: & $\varphi_m$  \\
static parameters: & $\varphi_r$, $\varphi_\text{PD}$, $z_{Rm}=z_{Rr}$, $z_m=z_r$, $x_{ir}$, $y_{ir}$ \\
variable parameters: & $x_{im}$, $y_{im}$ \\
detector: &  SEPD \\
\br
\end{tabular}

\begin{figure}
	\centering
	\includegraphics[width=0.9 \columnwidth]{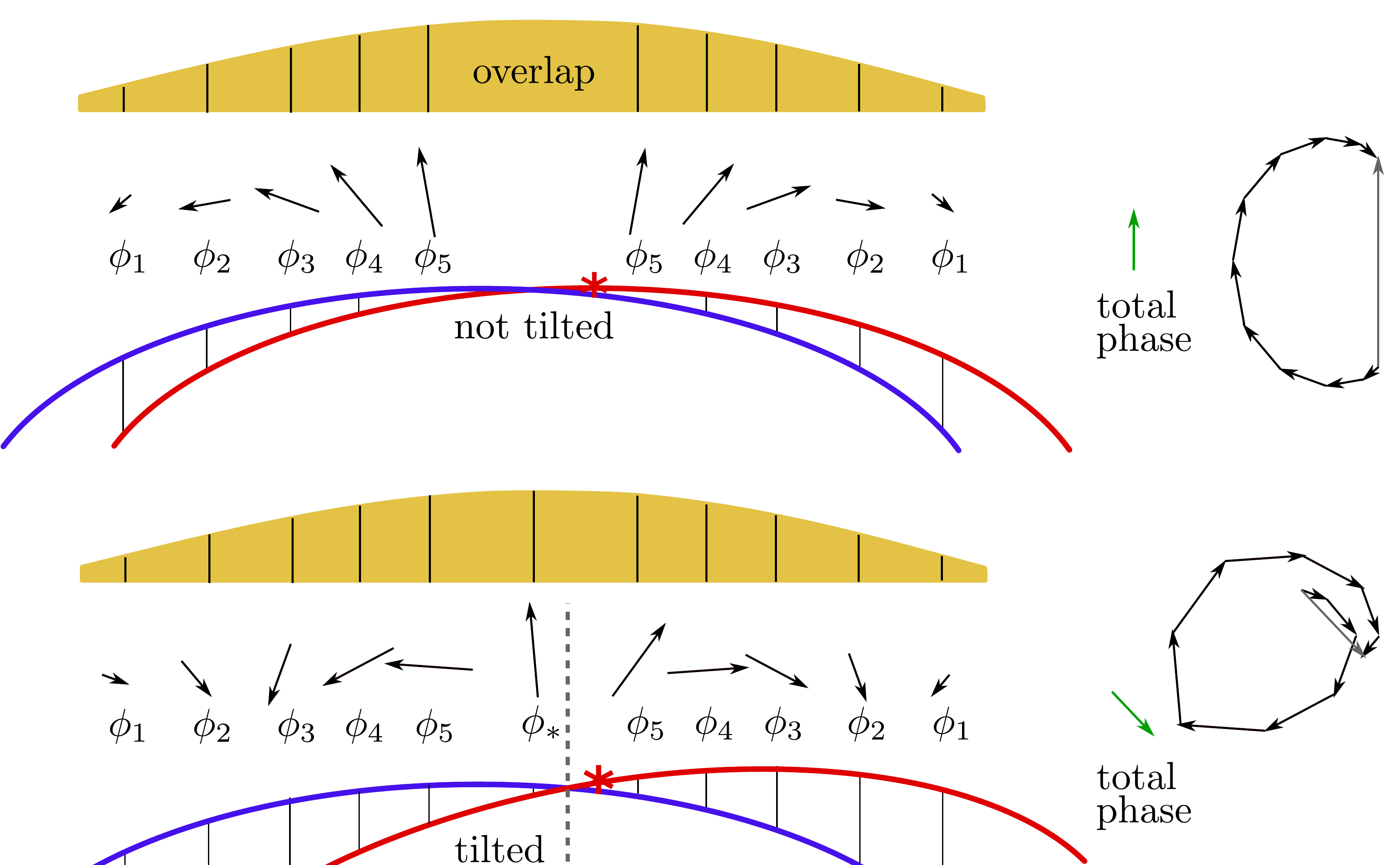}
	\caption{TTL coupling for a laterally shifted measurement beam (red), which rotates around its incidence point on the photodiode (red star). The phase information at different positions on the detector is weighted with the overlap (product of the electric field amplitudes) between the two interfering beams. 
	For non-tilted beams, the local phase differences cancel in the total phase (i.e.\ the green arrow points upwards).
	For a tilted measurement beam, this is no longer the case (i.e.\ the green arrow is rotated): the pairs of local phase differences $\phi_i$, $ i \in \{1,...,5\}$ distributed around the symmetry axis of the amplitude profile do not cancel each other. Consequently, the total phase changes, i.e.\ the sum of the paired local phase differences $\phi_i$ and the non-zero phase difference at the symmetry point of the amplitude profile at the photodiode surface $\phi_*$ is different than in the non-tilted case.
	This is caused by a displacement of the symmetry axis (dashed grey line) of the differential wavefront with respect to the symmetry axis of the amplitude profile of the total field. The differential wavefront describes the phase distribution at the detector surface, not considering the weighting by the amplitude. It is indicated by the vertical lines between the blue and red wavefront.
}
	\label{fig: beam walk 2}
\end{figure}%
By assuming a perfect overlap of both interfering beams for $\varphi_m=0$ as well as a centre of rotation positioned exactly at the point of incidence, we considered a very special case in Sec.~\ref{sec: non_geom equal}. 
We now relax these conditions step-wise. Within this subsection, we investigate how the TTL coupling changes if the two beams are laterally, i.e.\ along the $x$- or $y$-axis (Fig.~\ref{fig:PD_coordsyst}), shifted with respect to each other.
Let $x_{im},x_{ir}$ describe the offset of the beams' points of incidence with respect to a hypothetical centre of the infinitely large SEPD (see Fig.~\ref{fig:PD_coordsyst} for the coordinate system).
The measurement beam's centre of rotation is now placed in this new point, i.e.\ it jitters angularly around the point defined by $x_{im}$.

While without beam offsets, the amplitude of the overlap is symmetric with respect to the wavefronts,
the symmetry axes of the overlap and the interfering beams differ from each other once offsets of the detection points are introduced.
An offset of one beam and the resulting imbalance of the overlap will favour one side and thus generate cross-coupling, as demonstrated in Fig.~\ref{fig: beam walk 2}. 

We compute this case again analytically as described in \ref{sec: non_geom equal} for a statically tilted reference beam and photodiode and find
\begin{eqnarray}
\text{LPS}^\text{SEPD,2D}_\text{ng} %
&\approx  \frac{1}{2}(x_{im}-x_{ir})  \left[(\varphi_m-\varphi_\text{PD})+(\varphi_r-\varphi_{\text{PD}})\right] \nonumber\\
&\, + \left[ (x_{im}-x_{ir})^2\frac{z_m}{8 z_{Rm}^2} - \frac{z_m}{4k\,z_{Rm}} \right]  \left(\varphi_m^2-2\varphi_m\varphi_\text{PD}\right)   
\label{eq: non-geom equal offset long}
\end{eqnarray} 
or in a three-dimensional case neglecting detector angles
\begin{eqnarray}
\text{LPS}^\text{SEPD,3D}_\text{ng} &\approx 
 \frac{1}{2}(x_{im}-x_{ir})(\varphi_m+\varphi_r) \nonumber\\
&\,-\frac{1}{2}(y_{im}-y_{ir})(\eta_m+\eta_r) \nonumber\\
&\,+ \left[ (x_{im}-x_{ir})^2\frac{z_m}{8z_{Rm}^2}-\frac{z_m}{4k\,z_{Rm}} \right] \varphi_m^2 \nonumber\\
&\,+ \left[ (y_{im}-y_{ir})^2\frac{z_m}{8z_{Rm}^2}-\frac{z_m}{4k\,z_{Rm}} \right] \eta_m^2 \nonumber\\
&\,-\frac{z_m}{z_{Rm}^2+z_m^2}\,x_{im}y_{im}\,\varphi_m\eta_m \,. 
\label{eq:non_geom_3D_beam long}
\end{eqnarray}
Here, $y_{im},y_{ir}$ denote the vertical displacement of the points of incidence with respect to the hypothetical centre of the photodiode, and $\eta_m$ the measurement beam's pitch angle.
In both equations, we neglected all constant second-order angles (e.g.\ $(\varphi_r-\varphi_\text{PD})^2$ and $\varphi_\text{PD}^2$) but kept all linear constants in preparation for the following discussion. 

We can simplify Eqs.~\eqref{eq: non-geom equal offset long} and~\eqref{eq:non_geom_3D_beam long} further by discarding the negligible terms:
As shown in the discussion in Sec.~\ref{sec: non_geom equal}, the quadratic terms 
which include a division by the wavenumber $k$ are negligibly small for common setup parameters. 
Also, the offsets between the points of incidence are small in typical interferometers ($x_{im}-x_{ir}<10^{-4}$). 
Therefore, we further neglect the terms with quadratic offsets multiplied with quadratic beam tilts as fourth-order terms.
Correspondingly, the Eqs.~\eqref{eq: non-geom equal offset long} and~\eqref{eq:non_geom_3D_beam long} reduce to
\begin{eqnarray}
\text{LPS}^\text{SEPD,2D}_\text{ng} %
&\approx  \frac{1}{2}(x_{im}-x_{ir}) \left[(\varphi_m-\varphi_\text{PD})+(\varphi_r-\varphi_{\text{PD}})\right]  \,,
\label{eq: non-geom equal offset} \\
\text{LPS}^\text{SEPD,3D}_\text{ng} 
&\approx \frac{1}{2}(x_{im}-x_{ir})(\varphi_m+\varphi_r) \nonumber\\
&\,-\frac{1}{2}(y_{im}-y_{ir})(\eta_m+\eta_r) \,.
\label{eq:non_geom_3D_beam}
\end{eqnarray}
\begin{figure}
	\centering
	\includegraphics[width=0.9 \columnwidth]{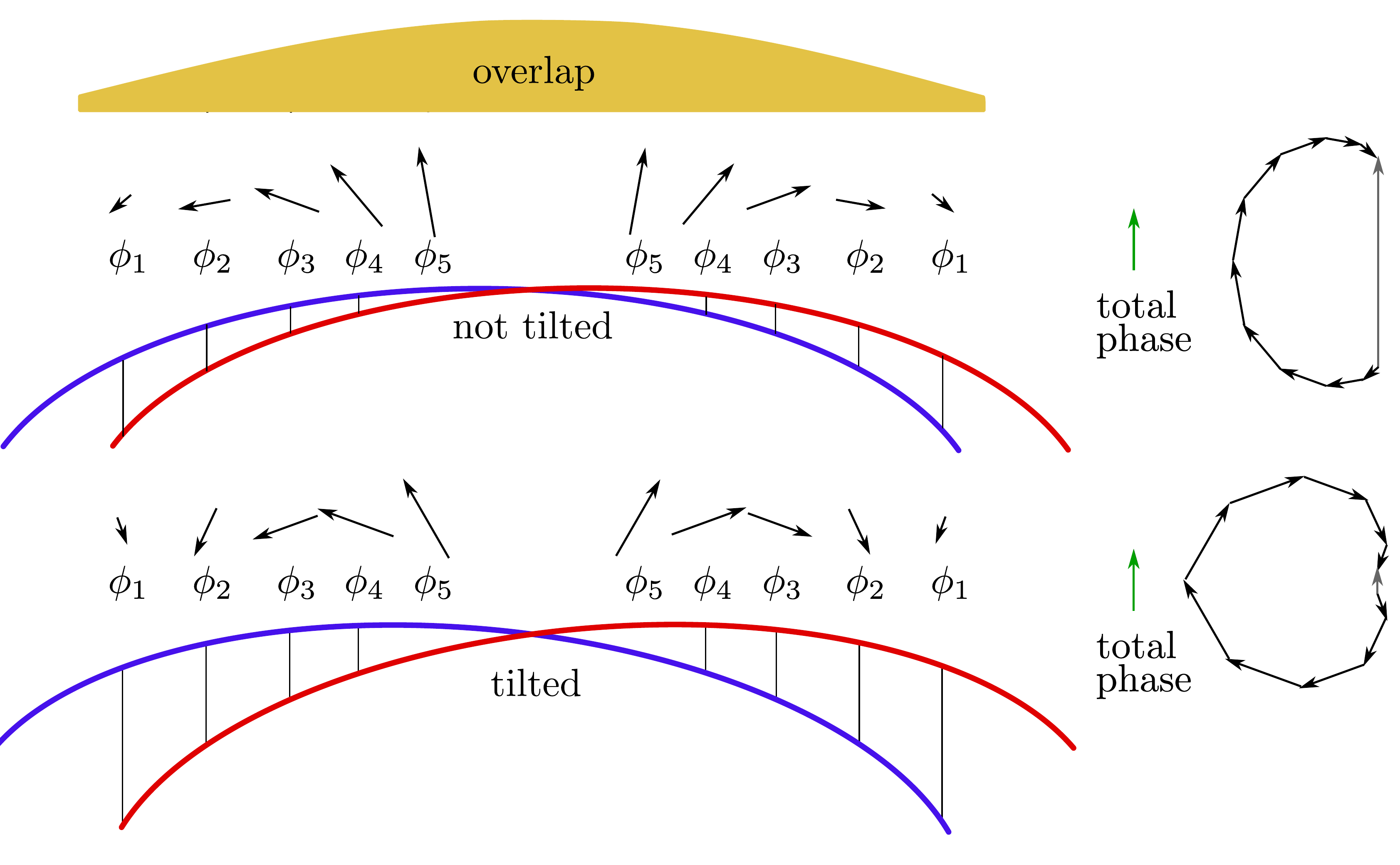}
	\caption{For two beams that are shifted with respect to each other and are rotated by the same angle but in opposite directions, i.e.\ $\varphi_m=-\varphi_r$, the amplitude profile is symmetric and the local phase differences cancel each other.}
	\label{fig: beam walk phi_r}
\end{figure}
It follows from Eq.~\eqref{eq: non-geom equal offset} that %
the TTL coupling would be minimised if the measurement and the reference beam are nominally tilted by inverse angles, i.e.\ $\varphi_m=-\varphi_r$, and $\varphi_\text{PD}=0$. 
The cancellation of the signal in the case of inverse angles is illustrated in Fig.~\ref{fig: beam walk phi_r}. This shows that not the differential but the average alignment angle of the beams couples into the signal.

Finally, we distinguish between a static and a dynamic point of incidence of the measurement beam.
In cases where the pivot is not actually on the detector, a beam walk occurs on the photodiode. That means, the incidence point $(x_\text{im}, y_\text{im})$ varies dynamically during the rotation, i.e.\ it is angle-dependent:
\begin{eqnarray}
  x_{im}(\varphi_m) &\approx x_{im,0} + x_{im}^\prime(0)\,\varphi_m + x_{im}^{\prime\prime}(0)\, \frac{\varphi_m^2}{2}
\label{eq:xim2D}
\end{eqnarray}
In our computations, we assume then a rotation of the measurement around this point of incidence.
We discuss the beam walk for different applications in Sec.~\ref{sec:BeamWalkMechanisms}.

In three-dimensional investigations, the point of incidence would depend on both jittering angles:
$(x_{im},y_{im}) = (x_{im}(\varphi_m,\eta_m),y_{im}(\varphi_m,\eta_m))$.
However, in cases where the beam walk can be linearised, no cross-plane dependencies appear.
This is the case for interfering beams with equal beam parameters and large single element diodes, as discussed in this section.
The point of incidence is then described by
\begin{eqnarray}
  x_{im}(\varphi_m,\eta_m) &\approx x_{im,0} + x_{im}^\prime(0)\,\varphi_m \,, 
\label{eq:xim3D}\\
  y_{im}(\varphi_m,\eta_m) &\approx y_{im,0} + y_{im}^\prime(0)\,\eta_m \,.
\label{eq:yim3D}
\end{eqnarray}

We insert for two-dimensional analyses Eq.~\eqref{eq:xim2D} into Eq.~\eqref{eq: non-geom equal offset} and for three-dimensional computation Eqs.~\eqref{eq:xim3D} and \eqref{eq:yim3D} into Eq.~\eqref{eq:non_geom_3D_beam} (where we set the photodiode alignment angle to zero).
When discarding all constant terms, we find
\begin{eqnarray}
  \text{LPF}_\text{ng}^\text{SEPD,2D} &\approx
  \frac{1}{2}(x_{im,0}-x_{ir})\,\varphi_m \nonumber\\
  &\,+ \frac{x_{im}^\prime(0)}{2}\,\varphi_m
  \left[(\varphi_m-\varphi_\text{PD})+(\varphi_r-\varphi_{\text{PD}})\right]  \,,
\label{eq:non_geom_2D_beam_dynamic} \\   
  \text{LPF}_\text{ng}^\text{SEPD,3D} &\approx
  \frac{1}{2}(x_{im,0}-x_{ir})\,\varphi_m + \frac{x_{im}^\prime(0)}{2}\,\varphi_m (\varphi_m+\varphi_r) \nonumber\\
  &\,-\frac{1}{2}(y_{im,0}-y_{ir})\,\eta_m 
  - \frac{y_{im}^\prime(0)}{2}\,\eta_m (\eta_m+\eta_r) \,.
\label{eq:non_geom_3D_beam_dynamic} 
\end{eqnarray}
We see that the constant beam offsets induce significant linear TTL coupling for angular jitter, while the dynamic beam walk term is purely quadratic in this case.

\begin{figure}
	\centering
	\includegraphics[width=0.85\columnwidth]{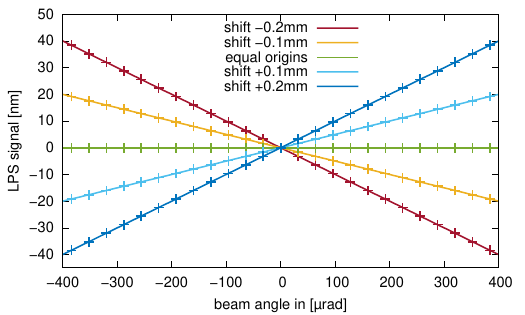}
	\caption{\label{fig: ttl with varying xim}Simulated path length signal (IfoCAD:\ crosses, analytical:\ line) in the scenario with two identical Gaussian beams that only differ in their nominal points of incidence at the detector $x_{im}$. This describes a shift of the axis of the measurement beam by $x_{im}$, while having the centre of rotation at the same longitudinal distance from the beam's point of incidence. 
	We find significant linear TTL coupling.
	The simulation parameters were: waist radius $w_0=1\,$mm, nominal points of incidence $x_{ir}=0$ and $x_{im}=\lbrace$\,-0.2\,mm, -0.1\,mm, 0\,mm, 0.1\,mm, 0.2\,mm$\rbrace$, centre of rotation at point of incidence, and detector radius 100\,mm.}
\end{figure}

If the point of incidence is static, i.e.\ do not change with the jitter, the Eqs.~\eqref{eq:non_geom_2D_beam_dynamic} and \eqref{eq:non_geom_3D_beam_dynamic} further reduce to
\begin{eqnarray}
\text{LPF}_\text{ng}^\text{SEPD,2D} &\approx
  \frac{1}{2}(x_{im}-x_{ir})\,\varphi_m \,,
\label{eq:non_geom_2D_beam_const_ximyim} \\
\text{LPS}^\text{SEPD,3D}_\text{ng} &\approx 
 \frac{1}{2}(x_{im}-x_{ir})\,\varphi_m -\frac{1}{2}(y_{im}-y_{ir})\,\eta_m \,. 
\label{eq:non_geom_3D_beam_const_ximyim}
\end{eqnarray}
This case is illustrated in Fig.~\ref{fig: ttl with varying xim}.

\subsection{Identical Gaussian beams and lateral jitter}
\label{sec: non_geom offset equal beam lateral}
\begin{tabular}{|ll|}
\br
jitter type: & $x_m$, $y_m$ \\
static parameters: & $\varphi_m$, $\varphi_r$, $\varphi_\text{PD}$, $z_{Rm}=z_{Rr}$, $z_m=z_r$, $x_{ir}$, $y_{ir}$ \\
variable parameters: & $x_{im}$, $y_{im}$ \\
detector: &  SEPD \\
\br
\end{tabular}

Next, we investigate translational beam jitter along the $x$- or $y$-axes (i.e.\ not the $z$-axis) of the detector (Fig.~\ref{fig:PD_coordsyst}).
We refer to this jitter as lateral jitter.
The lateral measurement beam jitter is related to the TTL coupling by the lateral beam offsets presented in the previous section.
This jitter would make the beam's point of incidence shift along the detector surface with respect to the reference beam's incidence point. 
Meanwhile, we assume its angular alignment ($\varphi_m$) to be constant.
It follows that the Eqs.~\eqref{eq: non-geom equal offset} and \eqref{eq:non_geom_3D_beam} still hold but must be interpreted for a variable point but constant angle of incidence of the measurement beam.
\begin{eqnarray}
  \text{LPS}^\text{SEPD,2D}_\text{ng} %
  &\approx  \frac{1}{2}\, x_m\, \left[(\varphi_m-\varphi_\text{PD})+(\varphi_r-\varphi_{\text{PD}})\right] \,,
\label{eq:non-geom_2D_equal_lateral} \\
  \text{LPS}^\text{SEPD,3D}_\text{ng} 
  &\approx \frac{1}{2}\, x_m\, (\varphi_m+\varphi_r)
  - \frac{1}{2}\, y_m\, (\eta_m+\eta_r) \,.
\label{eq:non_geom_3D_equal_lateral}
\end{eqnarray}
We see that lateral jitter would not couple if both beams are inversely tilted (compare Figure~\ref{fig: beam walk phi_r}).
However, for a differential angle $\varphi_m-\varphi_r\neq 0$, we find significant linear TTL coupling.

\subsection{Arbitrary Gaussian beams with arbitrary centre of rotation}
\label{sec: non_geom beam parameter}
\begin{tabular}{|ll|}
\br
jitter type: & $\varphi_m$ \\
static parameters: & $\varphi_r$, $\varphi_\text{PD}$, $z_{Rm},\, z_{Rr}$, $z_m,\, z_r$, $x_{ir}$ \\
variable parameters: & $x_{im}$ \\
detector: &  SEPD \\
\br
\end{tabular}

\begin{figure}
	\centering
	\includegraphics[width=0.85\columnwidth]{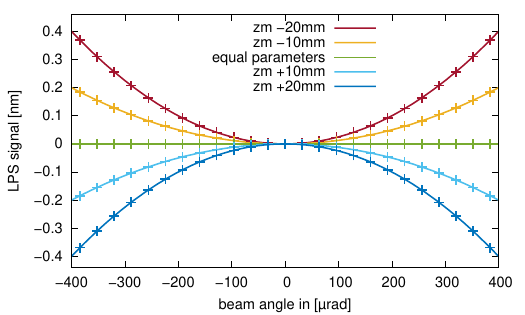}
	\caption{\label{fig: ttl with varying zm}Simulated path length signal (IfoCAD:\ crosses, analytical:\ line) in the scenario with two Gaussian beams that only differ in their waist location and interfere on an infinitely large detector. All graphs go through the origin because we chose for each setting the signal obtained at a beam angle of zero as a reference and subtracted it from each curve. 
	We find quadratic TTL coupling for this case of varying the waist location. 
	The simulation parameters were: waist radius of both beams $w_0=1\,$mm, nominal distances from waist at detector $z_r=100\,$mm and $z_m=z_r + \lbrace$\,-20\,mm, -10\,mm, 0\,mm, 10\,mm, 20\,mm$\rbrace$, pivot position at point of incidence, and detector radius 100\,mm. 
	}
\end{figure}
\begin{figure}
	\centering
	\includegraphics[width=0.85\columnwidth]{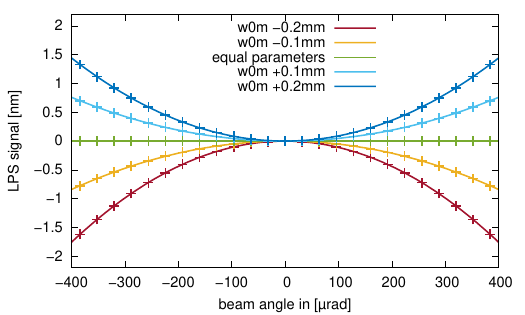}
	\caption{\label{fig: ttl with varying z0m} Simulated path length signal (IfoCAD:\ crosses, analytical:\ line) in the scenario with two Gaussian beams that only differ in their waist size and interfere on an infinitely large detector. All graphs go through the origin because we chose for each setting the signal obtained at a beam angle of zero as reference and subtracted it from each curve. 
	We find quadratic TTL coupling for this case of varying waist sizes. 
	The simulation parameters were: waist radii $w_{0r}=1\,$mm and $w_{0m}=w_{0r} + \lbrace$\,-0.2\,mm, -0.1\,mm, 0\,mm, 0.1\,mm, 0.2\,mm$\rbrace$, nominal distances from waist at detector $z_m=z_r=100\,$mm, $x_{im}=x_{ir}$, pivot position at point of incidence, and detector radius 100\,mm. 
	}
\end{figure}
In the previous examples, the measurement and reference beam initially had the same intensity and phase profile on the detector. However, this is not representative for interferometers like LISA, GRACE-FO, or arbitrary instruments that use heterodyne interferometers where identical beam parameters are effectively not achievable. 
A beam parameter mismatch has a strong influence on the cross-coupling.
The main reason for this is a discrepancy between the wavefront curvatures and the diameter of the beams in the detector plane. A qualitative analysis of this effect can be found in~\cite{schuster2016experimental}.

Here, we provide an analytic expression for the non-geometric part of the path length measurement as we have done for the previous simpler cases. We find
\begin{eqnarray}
 \text{LPS}^\text{SEPD,2D}_\text{ng}  &\approx
   (x_{im} - x_{ir}) \Bigg[ \frac{z_{Rm} (z_{Rm}+z_{Rr})+z_{m} (z_{m}-z_{r})}{(z_{Rm}+z_{Rr})^2+(z_{m}-z_{r})^2}\, (\varphi_m-\varphi_\text{PD}) \nonumber\\
   &\hspace*{2.3cm} +\frac{z_{Rr} (z_{Rm}+z_{Rr})-z_{r} (z_{m}-z_{r})}{(z_{Rm}+z_{Rr})^2+(z_m-z_r)^2}\,(\varphi_r-\varphi_\text{PD})\Bigg]
   \nonumber\\
  &-  
  \Bigg\lbrace\frac{z_{Rr} z_{m}+z_{Rm} z_{r}}{k((z_{Rm}+z_{Rr})^2+(z_{m}-z_{r})^2)} \nonumber\\
  &\quad + \frac{\left(x_{im} - x_{ir}\right)^2}{(z_{Rm}+z_{Rr})^2+(z_{m}-z_{r})^2} \left[z_{m}- \frac{2 (z_{Rm}+z_{Rr}) (z_{Rr} z_{m}+z_{Rm} z_{r})}{(z_{Rm}+z_{Rr})^2+(z_{m}-z_{r})^2} \right] \Bigg\rbrace \nonumber\\
  &\quad \cdot\left(\frac{\varphi_m^2}{2}-\varphi_m\varphi_\text{PD}\right)
  \nonumber\\
  &- 
  \left[  \frac{(z_{Rr}^2+z_r^2) z_{m}-(z_{Rm}^2+z_m^2) z_r}{(z_{Rm}+z_{Rr})^2+(z_{m}-z_{r})^2}\right] \left(\frac{\varphi_m^2}{2}-\varphi_m\varphi_r\right) \nonumber\\
  &+
   (x_{im}-x_{ir})^2\, \left[\frac{(z_m-z_r)}{2\left( (z_{Rm}+z_{Rr})^2+(z_m-z_r)^2\right)} \right] \,,
  \label{eq: non-geom unequal long}
\end{eqnarray} 
where $x_{im}$ can be either static or dynamic, as introduced in the previous subsection. 
Similar to the case of equal beam parameters, we can neglect some terms in Eq.~\eqref{eq: non-geom unequal long} since they are negligible in common interferometric setups. We neglected the terms featuring a division by the wavenumber $k$, or a product of a squared beam offset on the detector and a quadric tilt dependency, and find
\begin{eqnarray}
 \text{LPS}^\text{SEPD,2D}_\text{ng}  &\approx
   (x_{im} - x_{ir}) \Bigg[ \frac{z_{Rm} (z_{Rm}+z_{Rr})+z_{m} (z_{m}-z_{r})}{(z_{Rm}+z_{Rr})^2+(z_{m}-z_{r})^2}\, (\varphi_m-\varphi_\text{PD}) \nonumber\\
   &\hspace*{2.3cm} +  \frac{z_{Rr} (z_{Rm}+z_{Rr})-z_{r} (z_{m}-z_{r})}{(z_{Rm}+z_{Rr})^2+(z_m-z_r)^2}\,(\varphi_r-\varphi_\text{PD})\Bigg] 
   \nonumber\\
  &- 
  \left[  \frac{(z_{Rr}^2+z_r^2) z_{m}-(z_{Rm}^2+z_m^2) z_r}{(z_{Rm}+z_{Rr})^2+(z_{m}-z_{r})^2}\right] \left(\frac{\varphi_m^2}{2}-\varphi_m\varphi_r\right)  \nonumber\\
  &+ 
   (x_{im}-x_{ir})^2\, \left[\frac{(z_m-z_r)}{2\left( (z_{Rm}+z_{Rr})^2+(z_m-z_r)^2\right)} \right] \,.
  \label{eq: non-geom unequal}
\end{eqnarray}
By setting $z_{Rm}=z_{Rr}$ and $z_m=z_r$, Eq.~\eqref{eq: non-geom unequal} reduces to the case of equal beam parameters (compare Eq.~\eqref{eq: non-geom equal offset}).

As we have done before, we evaluate Eq.~\eqref{eq: non-geom unequal} for a beam walk of the measurement beam by inserting Eq.~\eqref{eq:xim2D}, which yields
\begin{eqnarray}
 \text{LPS}^\text{SEPD,2D}_\text{ng}  &\approx
   (x_{im,0} - x_{ir})\, \frac{z_{Rm} (z_{Rm}+z_{Rr})+[z_{m}+ x_{im}^\prime(0)] (z_{m}-z_{r})}{(z_{Rm}+z_{Rr})^2+(z_{m}-z_{r})^2}\, \varphi_m \nonumber\\
   &+ x_{im}^\prime(0)\,\varphi_m \Bigg[ \frac{z_{Rm} (z_{Rm}+z_{Rr})+z_{m} (z_{m}-z_{r})}{(z_{Rm}+z_{Rr})^2+(z_{m}-z_{r})^2}\, (\varphi_m-\varphi_\text{PD}) \nonumber\\
   &\hspace*{2.3cm} +  \frac{z_{Rr} (z_{Rm}+z_{Rr})-z_{r} (z_{m}-z_{r})}{(z_{Rm}+z_{Rr})^2+(z_m-z_r)^2}\,(\varphi_r-\varphi_\text{PD})\Bigg] 
   \nonumber\\
  &- 
  \left[  \frac{(z_{Rr}^2+z_r^2) z_{m}-(z_{Rm}^2+z_m^2) z_r}{(z_{Rm}+z_{Rr})^2+(z_{m}-z_{r})^2}\right] \left(\frac{\varphi_m^2}{2}-\varphi_m\varphi_r\right)  \nonumber\\
  &+ 
   \left[(x_{im}^\prime(0))^2 + x_{im}^{\prime\prime}(x_{im,0}-x_{ir})\right] \frac{(z_m-z_r)}{(z_{Rm}+z_{Rr})^2+(z_m-z_r)^2}\, \frac{\varphi_m^2}{2} \,.
  \label{eq: non-geom unequal dynamic}
\end{eqnarray}
For all static points of incidence, this expression reduces to
\begin{eqnarray}
 \text{LPS}^\text{SEPD,2D}_\text{ng}  &\approx
   (x_{im,0} - x_{ir})\, \frac{z_{Rm} (z_{Rm}+z_{Rr})+ z_{m} (z_{m}-z_{r})}{(z_{Rm}+z_{Rr})^2+(z_{m}-z_{r})^2}\, \varphi_m \nonumber\\
  &- 
  \left[  \frac{(z_{Rr}^2+z_r^2) z_{m}-(z_{Rm}^2+z_m^2) z_r}{(z_{Rm}+z_{Rr})^2+(z_{m}-z_{r})^2}\right] \left(\frac{\varphi_m^2}{2}-\varphi_m\varphi_r\right) \,.
  \label{eq: non-geom unequal static}
\end{eqnarray}

Though Eqs.~\eqref{eq: non-geom unequal}--\eqref{eq: non-geom unequal static} are fairly complex, they give valuable information, for instance, if only single parameter changes are investigated at a time.
Changing, for example, the difference in the distance from waist of the measurement and the reference beam gives a quadratic TTL coupling as shown in Fig.~\ref{fig: ttl with varying zm}.
Changing the waist sizes of the beam instead, while keeping the other parameters identical, will likewise generate second-order TTL coupling, see Fig.~\ref{fig: ttl with varying z0m}. This shows that the TTL coupling is a mixture of linear and second-order terms for arbitrary beam parameter mismatches.

\subsection{Arbitrary Gaussian beams and lateral jitter}
\label{sec: non_geom offset equal beam lateral}
\begin{tabular}{|ll|}
\br
jitter type: & $x_m$ \\
static parameters: & $\varphi_m$, $\varphi_r$, $\varphi_\text{PD}$, $z_{Rm}\,z_{Rr}$, $z_m,\,z_r$, $x_{ir}$ \\
variable parameters: & $x_{im}$ \\
detector: &  SEPD \\
\br
\end{tabular}

Analogously to the case with equal beam parameters, we obtain the equations for lateral jitter and arbitrary beams by interpreting the equations in the previous section for constant beam alignments and variable points of incidence:
\begin{eqnarray}
 \text{LPS}^\text{SEPD,2D}_\text{ng}  &\approx
   x_m \Bigg[ \frac{z_{Rm} (z_{Rm}+z_{Rr})+z_{m} (z_{m}-z_{r})}{(z_{Rm}+z_{Rr})^2+(z_{m}-z_{r})^2}\, (\varphi_m-\varphi_\text{PD}) \nonumber\\
   &\hspace*{2.3cm} +\frac{z_{Rr} (z_{Rm}+z_{Rr})-z_{r} (z_{m}-z_{r})}{(z_{Rm}+z_{Rr})^2+(z_m-z_r)^2}\,(\varphi_r-\varphi_\text{PD})\Bigg]
   \nonumber\\
  &+
   x_m^2\, \left[\frac{(z_m-z_r)}{2\left( (z_{Rm}+z_{Rr})^2+(z_m-z_r)^2\right)} \right] \,.
  \label{eq: non-geom unequal dynamic lateral}
\end{eqnarray}
Due to the different profiles of both beams, lateral jitter would now also yield a small coupling for non-tilted beams.

\subsection{Wavefront errors}
\label{sec: higher oder modes}
\begin{figure}
	\centering
	\includegraphics[width=0.5\linewidth , angle = {0}] {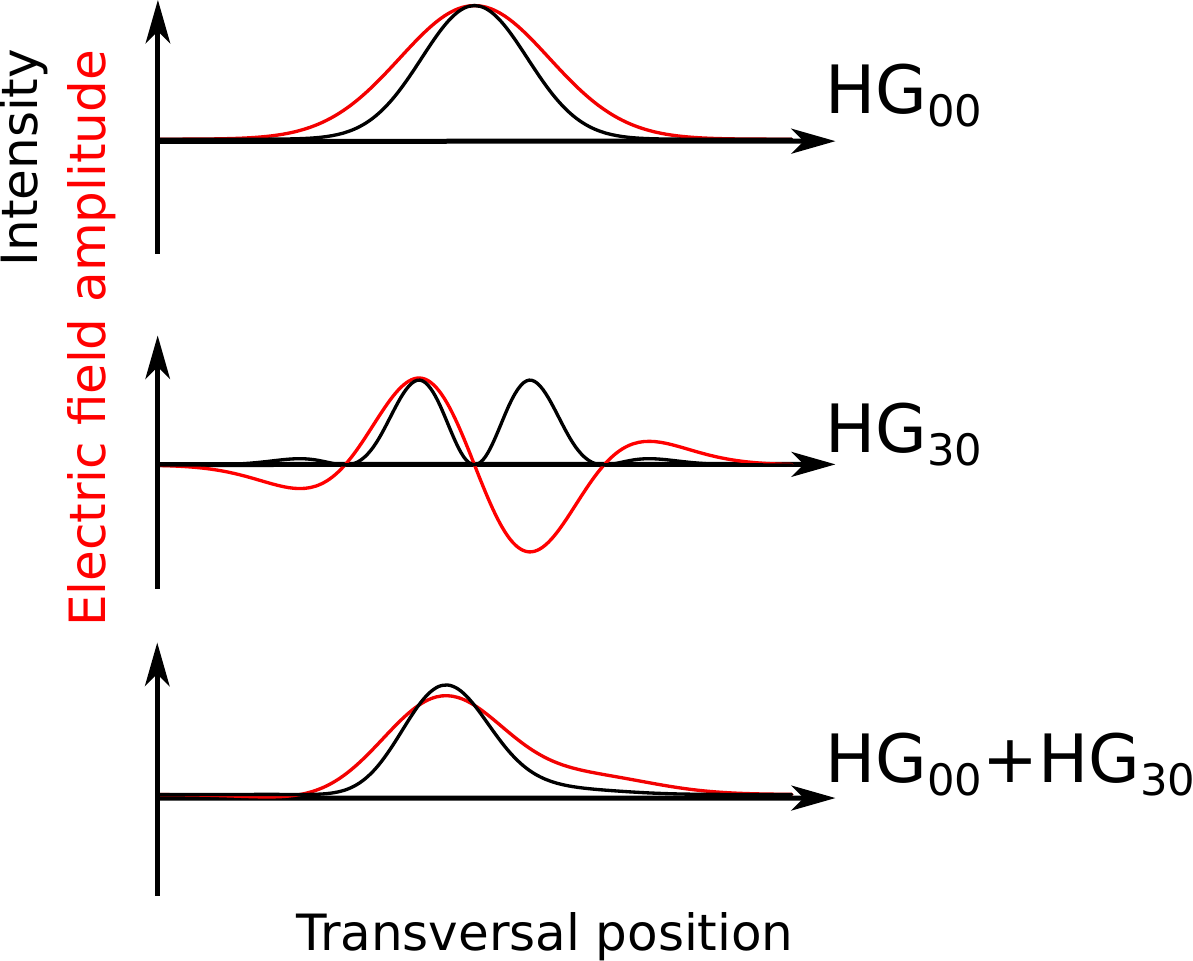}
	\caption{A fundamental mode is mixed with a HG$_\text{30}$ mode, the resulting amplitude and intensity profiles of the resulting beam are asymmetric (previously shown in \cite{schuster2016experimental}).}
	\label{fig: higher order modes}
\end{figure}

We have so far generally assumed the idealised case of perfect fundamental Gaussian beams with axial symmetry. In experimental reality, the interfering Gaussian wavefronts will have small distortions that can, for instance, be described by a superposition of higher-order modes with low amplitudes. These superimposing modes then affect the symmetry of the Gaussian beams, as illustrated in Fig.~\ref{fig: higher order modes}. Consequently, wavefront errors affect the TTL coupling behaviour. These additional wavefront error dependent TTL effects cannot be easily modelled for all the discussed cases. 
Instead, they produce very specific, non symmetric effects that must be
evaluated case-by-case. Thus we neglect them in the present discussion.
Some numerical simulation programs can handle these wavefront errors and consider them in the final signal, e.g., IfoCAD \cite{wanner2012methods,Kochkina2013}.

\section{TTL effects originating from detector properties}
\label{sec: photodiode geometry}
Not only do the wavefront properties of the beams affect the LPS signal, but also the detector geometry itself contributes to the cross-coupling. Mathematically, this can be seen from the integral over the detector surface $S$ in Eq.~\eqref{eq:ifo_phase}, which is used for computing an LPS signal using Eq.~\eqref{eq:LPS}.
We will therefore investigate below the various contributions to $\text{LPS}_\text{ng}$ originating from detector properties. 

In \cref{sec: photodiode angle}, we comment on the dependency of the LPS signal on the photodiode angle. We argue that it can be directly set to zero in simulations, independent of the corresponding experimental value. 
In \cref{sec: non geom signal definition}, we then show that different definitions of the LPS signal exist if quadrant diodes are used. These different LPS signal types usually show different amounts of TTL coupling. 
We then show in \cref{sec: detector size} how phase contributions for finite square quadrant photodiodes can be computed analytically. Finally, we briefly discuss the effect of diode imperfections in Sec.~\ref{sec:detector_errors}.

\subsection{Tilt of the detector}
\label{sec: photodiode angle}
\begin{tabular}{|ll|}
\br
jitter type: & ($\varphi_m$ or $x_m$) \\
static parameters: & (- or $\varphi_m$), $\varphi_r$, $\varphi_\text{PD}$, $z_{Rm}\, z_{Rr}$, $z_m\, z_r$, $x_{ir}$ \\
variable parameters: & $x_{im}$ \\
detector: &  SEPD \\
\br
\end{tabular}

\begin{figure*}
	\centering
	\includegraphics[width=0.98\textwidth]{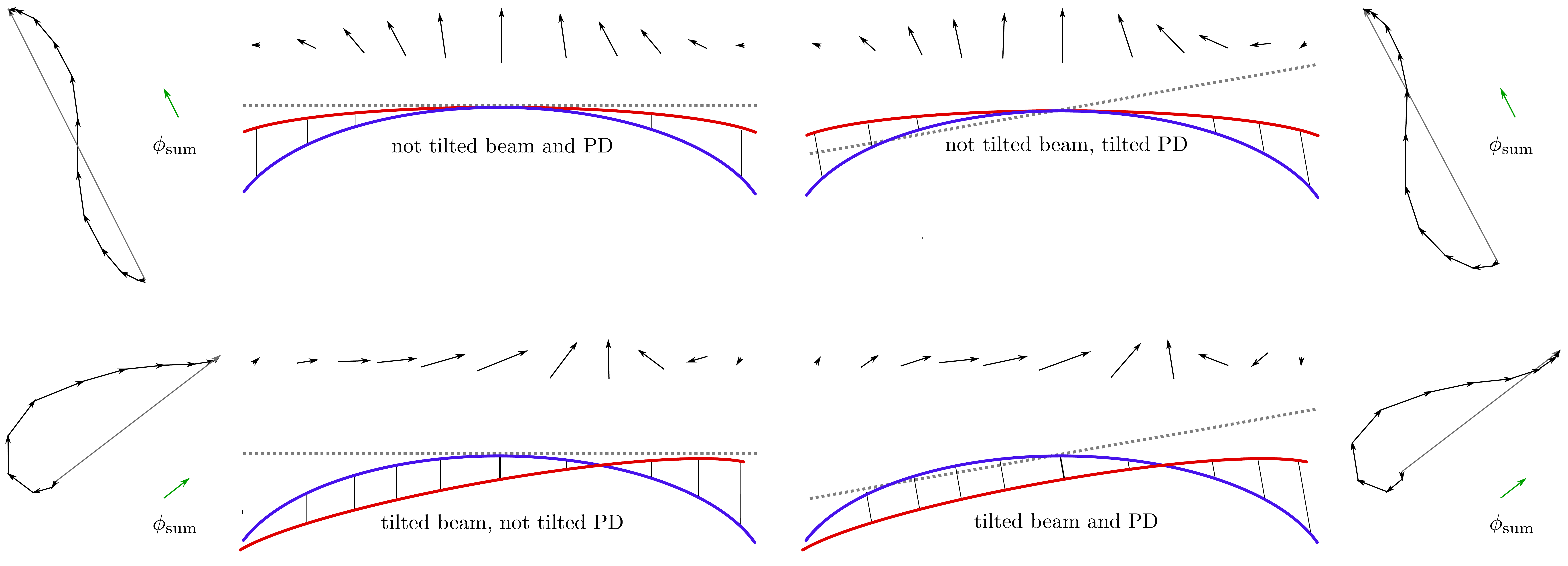}
	\caption{Qualitative analysis of the TTL coupling for unequal beam parameters showing that the photodiode angle does not contribute to the TTL coupling. The left-hand side shows the case of a non-tilted detector ($\varphi_\text{PD}=0$), and the right-hand side the case of a tilted detector. Upper left figure: Neither the beams nor the detector is tilted. Lower left figure: Non-tilted detector, but the measurement beam (red) got tilted with respect to the reference beam (blue). The total phase changed in comparison to the non-tilted case, implying TTL coupling. 
	Upper and lower right figures: same as the corresponding left figures, but the detector surface was tilted. This tilt changes the direction along which the complex amplitudes are being defined and read out (see connecting lines between the blue and red curves), such that all vector representations of the complex amplitudes are slightly changed. However, no change in the total phases can be observed here compared to the cases without detector tilt. 
}
	\label{fig: cancellation PD tilt}
\end{figure*}

The equations shown in the previous sections depend on the TTL coupling in the non-geometric LPS signal on the angular alignment of the detector surface ($\varphi_\text{PD}$).
While this is true for purely non-geometric coupling and also the geometric counterpart presented in \cite{G21},
this does not necessarily also hold for the full interferometric phase. 
For small tilts of the detector surface, it would in a good approximation
equally affect the reference and measurement beam, and likewise their relative phase in each detector point.

This is shown in Fig.~\ref{fig: cancellation PD tilt}.
There, the TTL coupling for unequal beam parameters is compared for the case of an un-tilted photodiode (left part of the image) with the case of a tilted photodiode (right part of the image). Even though the complex amplitude vectors in each detector point slightly change due to the detector tilt, we do not see an effect on the total phase (green arrows).
Hence, the detector angle is expected to cancel from the LPS signal. 

We confirm this expectation by analysing the photodiode angle dependent terms in $\text{LPS} = \text{OPD} + \text{LPS}_\text{ng}$. 
Therefore, we subtract all terms not depending on $\varphi_\text{PD}$, i.e.\ LPS($\varphi_\text{PD}=0)$, and find 
\begin{eqnarray}
	 & \text{LPS}^\text{SEPD,2D}(\varphi_\text{PD}) - \text{LPS}^\text{SEPD,2D}(\varphi_\text{PD}=0) \nonumber \\
	 \approx  &
	 \text{OPD}^\text{SEPD,2D}(\varphi_\text{PD}) - \text{OPD}^\text{SEPD,2D}(\varphi_\text{PD}=0)
	 -2\varphi_\text{PD}\,\varphi_m\,x_{im}^\prime(0) \,
\label{eq: phi_pd term}
\end{eqnarray}
for angular jitter coupling.
We see that all considerable non-geometric detector tilt dependent TTL terms, except for those of the jitter dependent beam walk, cancel out in the full LPS signal.
Particularly, the detector alignment angle does not add up to the TTL coupling for static offsets of the points of incidence. 
We will further show in Sec.~\ref{sec:BeamWalkMechanisms} that the residual, beam walk dependent terms in Eq.~\eqref{eq: phi_pd term} cancel with their geometric counterparts in different setups. This yields
\begin{equation}
	 \text{LPS}^\text{SEPD,2D}(\varphi_\text{PD}) - \text{LPS}^\text{SEPD,2D}(\varphi_\text{PD}=0) \approx  0 \,.
\label{eq: phi_pd term zero}
\end{equation}

We have seen above that the equations for lateral jitter can be obtained by a different interpretation of the terms evaluated for angular jitter. 
If follows that the conclusions for angular jitter coupling also hold for lateral jitter.

We conclude from this that in any derivation of LPS signals for large SEPDs, the photodiode angle $\varphi_\text{PD}$ can be neglected even if the diode is in the corresponding experiment indeed tilted.
Nonetheless, we show all equations with $\varphi_\text{PD}$ here because ray tracing tools computing the OPD will always include this term. 
Since the results can yield misleading alignment optimisation strategies regarding the detector alignment, it can be advisable to set $\varphi_\text{PD}=0$ in simulations.

\subsection{Dependence on the path length signal definition using QPDs}
\label{sec: non geom signal definition}
So far, we have assumed infinitely large SEPDs as detectors, which means that both wavefronts are fully sensed by the detector and no clipping occurs. This assumption is not valid in cases where a quadrant photodiode (QPD) is being used for angular sensing using differential wavefront sensing (DWS) \cite{Morrison1994}.
QPDs further allow for multiple phase signal definitions \cite{wanner2015brief}, which contribute differently to the overall cross-coupling. This originates from the fact that every photodiode quadrant delivers an individual photocurrent and, therefore, an individual complex amplitude. Mathematically, this is described by evaluating the surface integral in Eq.~\eqref{eq:Def_ComplexAmp_i} over the corresponding quadrant $i$, resulting in the complex amplitude $a_i$.
The four complex amplitudes can then be combined in different ways to generate the total phase readout $\phi$ and the corresponding LPS signal.
We show this here exemplary for two commonly used QPD path length definitions, the arithmetic mean phase and the LISA Pathfinder LPS signal.

The arithmetic mean phase (AP) is literally derived from the arithmetic mean of the phases of the four segments 
\begin{eqnarray}
	\phi^\text{AP} & =\frac{\arg(a_\text{A}) + \arg(a_\text{B}) +\arg(a_\text{C}) + \arg(a_\text{D})}{4} \\
	& =\frac{\phi_\text{A} + \phi_\text{B} +\phi_\text{C} + \phi_\text{D}}{4} \,,
\label{eq: phase qpd ap}	  
\end{eqnarray}
where A,B,C,D denote the four quadrants of the photodiode.
The corresponding LPS signal is then defined like before (see Eq.~\eqref{eq:LPS}): 
\begin{equation}
	\mathrm{LPS}^\text{AP}  = \frac{\phi^\text{AP}}{k}
	 =\frac{\phi_\text{A} + \phi_\text{B} +\phi_\text{C} + \phi_\text{D}}{4k} \,.
\label{eq: signal qpd ap}	  
\end{equation}
This is a kind of natural procedure when the signal is processed by a digital phase-locked loop (DPLL) based phasemeter that produces phases as primary output and not complex amplitudes \cite{DPLL2020}.

The second signal definition we discuss here is called LISA Pathfinder (LPF) LPS signal definition because it was used in the LISA Pathfinder phasemeter, which uses a single bin discrete Fourier transform (SBDFT)~\cite{heinzel2003interferometry}. The result of which is a complex amplitude. It is defined as the argument of the sum of all complex amplitudes divided by $k$:
\begin{equation}
	\mathrm{LPS}^\text{LPF} = \frac{1}{k} \arg(a_\text{A} + a_\text{B} + a_\text{C} + a_\text{D})\,.
\end{equation}

For a QPD with slits of zero width, the sum of the complex amplitudes of the single quadrants, corresponding to a sum of the integrals of the single segments, is equal to the complex amplitude of the entire diode. Therefore, for a slit width of zero and a sufficiently large diameter of the diode, the LISA Pathfinder QPD path length definition ($\mathrm{LPS}^\text{LPF}$) becomes equal to the SEPD path length definition ($\mathrm{LPS}^\text{SEPD}$)~\cite{wanner2015brief}.

\begin{figure}
	\centering
	\includegraphics[angle=0,width= \linewidth]{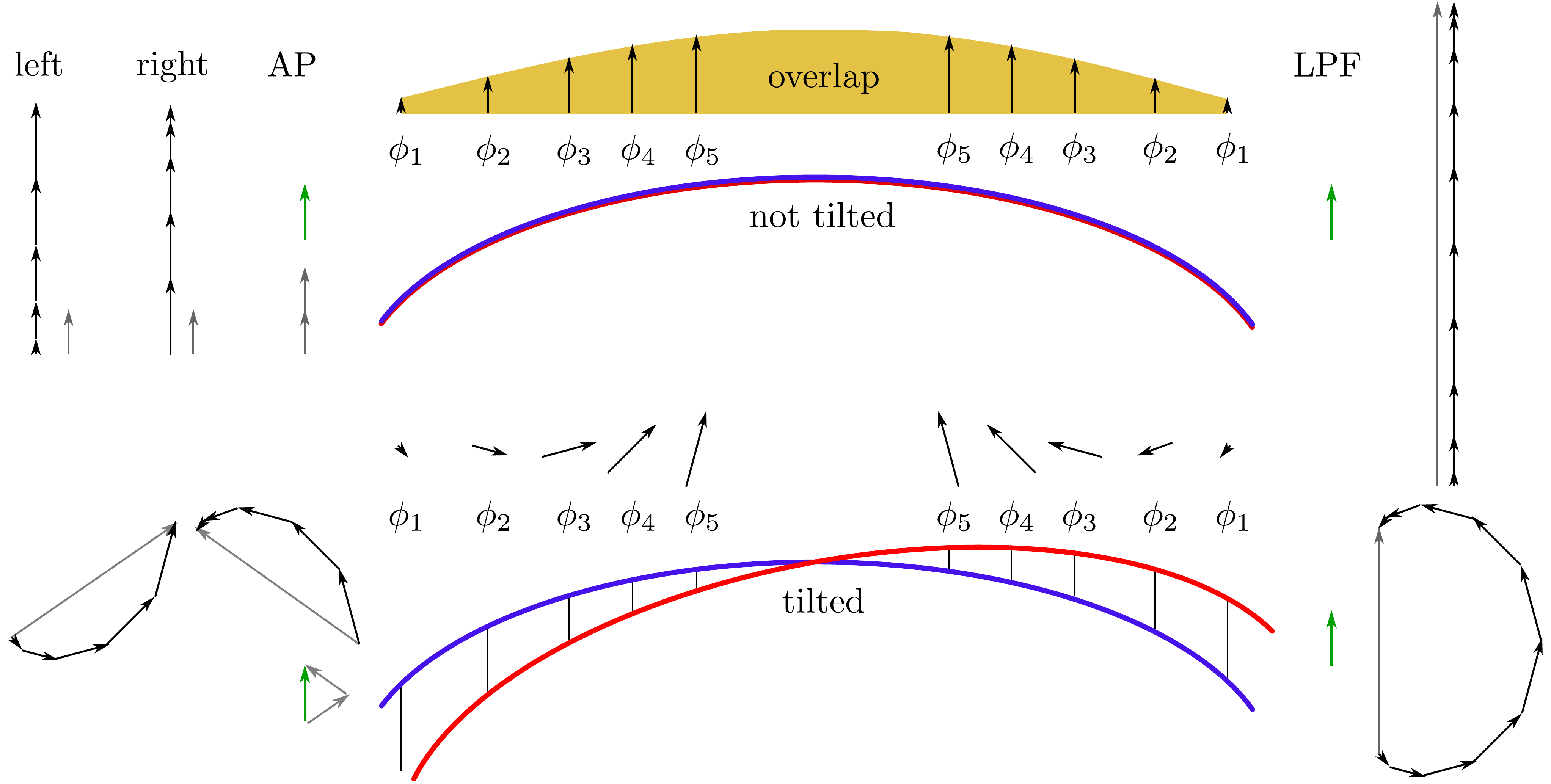}
	\caption{\label{fig: wavefront equal}The effect of different path length definitions on the overall phase in the case of equal wavefront curvatures.
	On the right, the LPF definition sums up all complex amplitude vectors.
	On the left, the AP definition, i.e.\ Eq.~\eqref{eq: signal qpd ap}, is used for two segments, which computes an average phase per side (left and right) and adds the averaged side complex amplitude vectors (grey arrows). The overall phase (green arrow) is independent of the tilt angle for both phase definitions. } 

	\centering
	\includegraphics[angle=0,width=\linewidth]{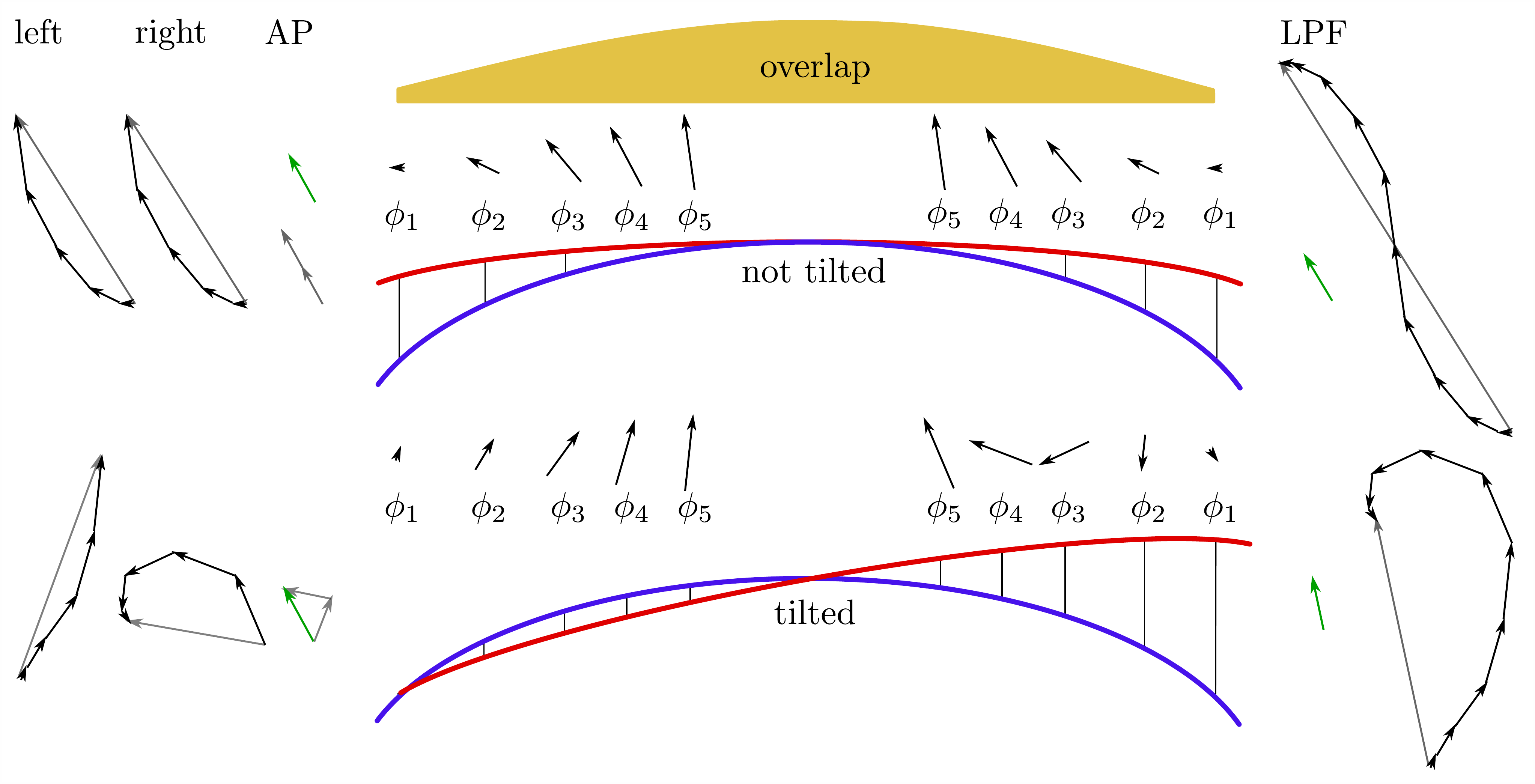}
	\caption{\label{fig: wavefront unequal} The effect of different path length definitions on the overall phase in case of unequal wavefront curvatures.
	On the right, the LPF path length signal is shown, and on the left, the AP definition. 
	Both, the LPF and the AP phases, change with the rotation. However, the AP phase is less affected by the tilts. 
	}
\end{figure}

Figures~\ref{fig: wavefront equal} and~\ref{fig: wavefront unequal} illustrate the effect of different wavefront curvatures on the cross-coupling if either the LPS$^\text{LPF}$ or LPS$^\text{AP}$ signal is used.
In the centre, two interfering wavefronts are shown, with equal curvatures in Fig.~\ref{fig: wavefront equal} and with unequal curvatures in Fig.~\ref{fig: wavefront unequal}.
Depending on the total phase signal definition, the complex amplitude vectors are recombined here in different ways. 
On the right, the LPF definition sums up all shown complex amplitude vectors of the left- and right-hand sides. 
On the left, the AP signal calculation is illustrated. Here, the complex amplitude vectors of the left and right-hand sides are added separately, resulting in a phase of the left- and right-hand sides (small grey arrows). These complex amplitude vectors are then added and normalised again, resulting in the green arrow, which illustrates the averaged phase.
In the case of equal wavefront curvatures, Fig.~\ref{fig: wavefront equal}, the overall phase (i.e.\ the angle of the green arrow) is independent of the tilt angle for both phase definitions. 
That means that, like in the case of an SEPD, we do not expect TTL coupling for either of the LPS definitions if the interfering wavefronts have matched beam parameters.

\begin{figure}
	\centering
	\includegraphics[width=0.85\columnwidth]{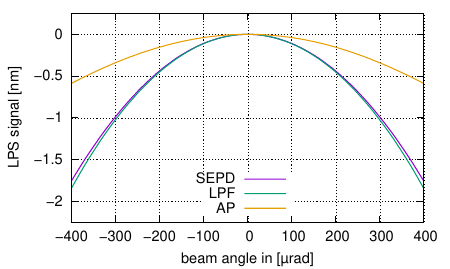}
	\caption{\label{fig:ifocad_signal_def} Longitudinal path length signals computed with IfoCAD for a single element diode, and for a QPD with either the LPF or the AP signal definition. Both beams are nominally equally aligned. One of the beams got tilted around its point of incidence, ensuring a pure non-geometric signal. The simulation parameters were: waist radius $w_{0r}=$1\,mm and $w_{0r}=$0.8\,mm, distance from waist at detector $z_m=z_r=100\,$mm, pivot position  $d_\text{long}=d_\text{lat}=0\,$mm, detector radius 5\,mm, and slit width 50\,$\mu$m.}
\end{figure}

Contrary, in the case of unequal wavefront curvatures in Fig.~\ref{fig: wavefront unequal}, the overall LPF phase changes if the wavefront tilts, while the overall AP phase is less affected by tilts. Hence, in the case of unmatched wavefront properties and a rotation around the centre of the QPD, we expect that the AP signal shows less TTL coupling than the LPF signal. 
We confirm this using IfoCAD as shown in Fig.~\ref{fig:ifocad_signal_def} for two aligned beams with identical parameters besides their waist size and a rotation of the measurement beam around their shared point of incidence. By Fig.~\ref{fig:ifocad_signal_def}, we demonstrate that the resulting AP signal comprises less TTL coupling than the LPF signal. Furthermore, the image shows that the LPF signal and the SEPD signal have nearly identical TTL coupling. This is expected from the definition of LPS$^\text{LPF}$. Finally, this demonstrates that using quadrant diodes and an AP signal can, in some cases, reduce the TTL coupling noise in the system. This was previously also observed in \cite{SchusterPhD, TN8.1}.

\subsection{TTL contributions comparing infinite SEPDs with finite square QPDs}
\label{sec: detector size}
We show in this section that the TTL estimates for large SEPDs are still useful in cases where small square SEPDs or QPDs and the LPF LPS signal are used. The derived signal contains the large SEPD's signal, though an additional TTL term originating from the clipping needs to be considered.

To account for phase changes due to clipping on the detector surface, we repeat the analytic derivation introduced in Sec.~\ref{sec: non geom analytic derivation} but assume this time a square finite quadrant photodiode shape. 
This means we integrate the complex product of the electrical fields over the detector surface, see Eq.~\eqref{eq:ifo_phase} and compare \cite{wanner2012methods, Wanner2014}.
Transforming the electrical fields into photodiode coordinates $x,y$, we can rewrite Eq.~\eqref{eq:ifo_phase} via 
\begin{eqnarray}
\phi = \arg\left[ \int_\text{PD}\mathrm{d}S\, \exp\left(C_{xx} x^2 + C_x x+C_{yy} y^2 + C_y y +C_0\right) \right]
\label{eq: phase pd coordinate system}
\end{eqnarray}
where $C_i\in\mathbb{C}$. 
The integration over the detector surface gives
\begin{eqnarray}
\phi &= \arg\Bigg\lbrace \frac{\pi \exp\left[\frac{1}{4}\left(\frac{C_x^2}{C_{xx}}+\frac{C_y^2}{C_{yy}}-4C_0\right)\right]}{\sqrt{C_{xx}}\sqrt{C_{yy}}} \nonumber\\
&\qquad \times \left[ \frac{1}{4}\,\mathrm{erf}\left( \frac{C_x+C_{xx}\, x}{2\sqrt{C_{xx}}} \right)\right]
\bigg[ \frac{1}{4}\,\mathrm{erf}\bigg( \frac{C_y+C_{yy}\, y}{2\sqrt{C_{yy}}} \bigg)\bigg] \bigg\vert_\text{PD}\Bigg\rbrace
\\
&= \arg\Bigg\lbrace \frac{\pi \exp\left[\frac{1}{4}\left(\frac{C_x^2}{C_{xx}}+\frac{C_y^2}{C_{yy}}-4C_0\right)\right]}{\sqrt{C_{xx}}\sqrt{C_{yy}}} \Bigg\rbrace \nonumber\\
&+ \arg\bigg\lbrace \bigg[ \frac{1}{4}\,\mathrm{erf}\bigg( \frac{C_x+C_{xx}\, x}{2\sqrt{C_{xx}}} \bigg)\bigg]
\bigg[ \frac{1}{4}\,\mathrm{erf}\bigg( \frac{C_y+C_{yy}\, y}{2\sqrt{C_{yy}}} \bigg)\bigg] \bigg\vert_\text{PD}\bigg\rbrace \,,
\label{eq: phi arbitrary pd shape}
\end{eqnarray}
whereby the notation $\int \mathrm{d}S\, f(x,y)= F(x,y)\vert_\text{PD}$ describes the evaluation of the antiderivative $F(x,y)$ over the surface of the photodiode.
If the photodiode surface consists of several segments, the notation describes the sum over these segments, inserting the boundary values respectively.

It can easily be verified that the second argument in Eq.~\eqref{eq: phi arbitrary pd shape} becomes zero for infinitely large SEPDs. Hence
\begin{eqnarray}
\phi &= \phi_\text{SEPD} \nonumber\\
&+ \arg\bigg\lbrace \left[ \frac{1}{4}\,\mathrm{erf}\left( \frac{C_x+C_{xx}\, x}{2\sqrt{C_{xx}}} \right)\right]
\left[ \frac{1}{4}\,\mathrm{erf}\left( \frac{C_y+C_{yy}\, y}{2\sqrt{C_{yy}}} \right)\right] \bigg\vert_\text{PD}\bigg\rbrace \, .
\label{eq: phi arbitrary pd shape SEPD dependent}
\end{eqnarray}
We conclude that the phase measured, for instance, by a QPD is always the sum of the phase measured by an infinitely large SEPD and a second phase contribution that depends on the detector geometry. However, for large diodes with small deviations from SEPDs, as for example QPDs with a narrow insensitive slit, $\phi_\text{SEPD}$ will be the dominant summand.
This agrees exactly with the simulation results shown in Fig.~\ref{fig:ifocad_signal_def}, where a QPD with a slit width of 50\,$\mu$m was assumed and the LPS$^\text{LPF}$ signal deviates only slightly from the SEPD signal.

\subsection{Arbitrary detector errors}
\label{sec:detector_errors} 
As for arbitrary wavefront errors, also imperfections of the detector affect the TTL coupling. The segments can have different efficiencies or different shapes due to defects or additional features like bonding wires.
Such disturbances are described and analysed for different detectors, for instance, in \cite{barranco2016spatially}.
In such a case, the equations described throughout this paper would be disturbed by the imperfections and would, therefore, need to be adapted to the given situation. 
This is not possible for arbitrary and unknown detector imperfections.

\section{Non-geometric TTL coupling in different systems}
\label{sec:BeamWalkMechanisms}
TTL coupling occurs in different types of precision interferometers. 
We group these interferometers into two categories, just as described in \cite{G21}. 
The first category covers systems, where the TTL coupling originates from the jitter of a system relative to a static beam. This occurs, for instance, if a spacecraft is jittering relative to the laser beam it is receiving from a far spacecraft, like in the GRACE-FO mission \cite{sheard2012intersatellite,GFOabich19} and in the LISA long-arm interferometers \cite{LISAMission,Jennrich2009}. 
The second category of interferometers comprises systems where the TTL coupling originates from an angularly or laterally jittering mirror. 
Most laboratory systems, as well as the LISA Pathfinder interferometers \cite{mcnamara2008lisa,armano2016sub,armano2019lpf} and LISA test mass interferometer \cite{LISAMission,Jennrich2009} fall into this category.

\subsection{Receiver jitter}
\label{sec:BeamWalkMechanisms_Receiver}

\begin{figure}
	\centering
	\includegraphics[width=0.7\columnwidth]{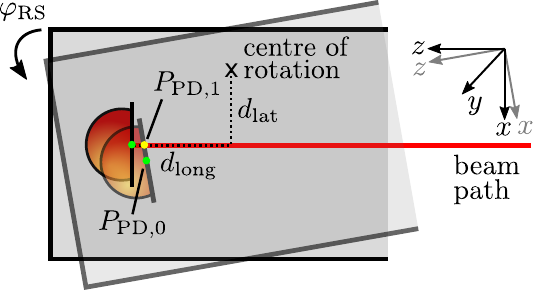}
	\caption{TTL coupling due to angular jitter of the system (grey open box) with respect to the incoming beam (red trace). The geometric TTL effect is visible by the distance change along the beam axis when the photodiode (PD) moves into the received beam. Non-geometric effects occur, for instance, due to the change of the beam's incidence point on the detector, which moves from the green point in the centre of the diode, to a yellow point $P_\text{PD,1}$ shifted from this centre.  
	 The distances $d_\text{long}$ and $d_\text{lat}$ define the longitudinal and lateral distances between the nominal point of incidence $P_\text{PD,0}$ and the centre of rotation. Both are positive in this figure.}
	\label{fig:ngeoSCTTL}
\end{figure}

We first assume an angularly jittering receiving system. There, a jittering optical bench receives a static beam which is then incident on a photodiode that jitters together with the optical bench and all other components, including the reference beam in the system.
Thus, the detector surface moves into or out of the received beam, which alters the optical path length and, thereby, introduces geometric TTL coupling.
Simultaneously it changes the wavefront properties on the photodiode and causes beam walk, which again results in non-geometric TTL coupling.
The beam walk is illustrated in Fig.~\ref{fig:ngeoSCTTL}, where $P_\text{PD,0}$ was the nominal point of incidence, while the beam impinges in point $P_\text{PD,1}$ if the system is rotated by $\varphi_\text{RS}$. 

We define the lateral jitter of a receiver as the jitter along the $x$- or $y$-axis of the receiver, see Fig.~\ref{fig:ngeoSCTTL}.
Due to this jitter, the measurement beam would get shifted along the detector surface and, therefore, the offset $x_{im}$ of the point of detection of the measurement beam with respect to the detector centre becomes time-dependent.

\subsubsection{Angularly jittering receiving system}
\label{sec: SR-arbitrary_pivot_identical_Gaussians}
\ \\
\begin{tabular}{|ll|}
\br
jitter type: & $\varphi_\text{RS}$, $\eta_\text{RS}$ \\
static parameters: & $d_\text{long}$, $d_\text{lat}$, $d_\text{vert}$ \\
 & $\varphi_r,\,\eta_r$, $\varphi_\text{PD}$, $z_{Rm}=z_{Rr}$, $z_m=z_r$, $x_{im,0},\,x_{ir}$, $y_{im,0},\,y_{ir}$ \\
detector: &  SEPD \\
\br
\end{tabular}

We start with investigating
the angular jitter of a receiving system. 
Here, the offset of the point of rotation from the nominal point of incidence causes a beam walk on the detector surface.
Thereby, the centre of rotation considered in the derivation of the non-geometric TTL signal becomes angle-dependent.

In the case of an angularly jittering receiver, the pivot point is shifted longitudinally by  $d_\text{long}$, laterally by $d_\text{lat}$, and vertically by $d_\text{vert}$ with respect to the point of incidence ($x_{im},{y_{im}}$).
The beam's horizontal and vertical offsets on the detector can be computed geometrically.
We find for two-dimensional investigations
\begin{eqnarray}
x_{im,\text{RS}}^\text{2D} &= \, x_{im,0} + \left\lbrace d_\text{long}\, \sin(\varphi_\text{RS}) - d_\text{lat}\, \left[ \sec(\varphi_\text{RS})-1\right]\right\rbrace \nonumber\\
&\hspace*{13mm} \cdot\cos(\varphi_\text{RS})\sec(\varphi_\text{RS}-\varphi_\text{PD})
\\
&\approx \, x_{im,0} +\, d_\text{long}\, \varphi_\text{RS} - \frac{1}{2}d_\text{lat}\, \varphi_\text{RS}^2 \ ,
\label{eq: x_im beam CoR}
\end{eqnarray} 
where $x_{im,0}$ is the measurement beam's initial offset, i.e.\ its offset at $\varphi_\text{RS}=0$. 

Extending our analysis to the three-dimensional case, we find for the beam walk (linearised)
\begin{eqnarray}
x_{im,\text{RS}}^\text{3D} &\approx x_{im,0} + d_\text{long}\,\varphi_\text{RS} \,, \\
y_{im,\text{RS}}^\text{3D} &\approx y_{im,0} - d_\text{long}\,\eta_\text{RS} \,.
\end{eqnarray}

We insert these beam walk equations into the non-geometric TTL coupling equations introduced in Sec.~\ref{sec: non_geom offset equal beam} for equal beam parameters, i.e.\ Eqs.~\eqref{eq:non_geom_2D_beam_dynamic} and \eqref{eq:non_geom_3D_beam_dynamic} (where $\varphi_\text{PD}=0$), respectively, and substitute $\varphi_m\rightarrow-\varphi_\text{RS}$ and $\eta_m\rightarrow\eta_\text{RS}$. 
Then, we get the non-geometric TTL contributions originating from the pivot location of the receiving system 
\begin{eqnarray}
\text{LPS}^\text{SEPD,2D}_\text{ng,RS} %
&\approx - \frac{1}{2}(x_{im,0}-x_{ir}) \varphi_\text{RS}  \nonumber\\
&+ \frac{1}{2}\, d_\text{long}\, \varphi_\text{RS} \left[-\varphi_\text{RS} + \varphi_r-2\varphi_{\text{PD}}\right] \,,
\label{eq: non-geom equal offset beam CoR} \\
\text{LPS}^\text{SEPD,3D}_\text{ng,RS} &\approx 
  -\frac{1}{2}(x_{im,0}-x_{ir}) \varphi_\text{RS} 
  +\frac{1}{2}(y_{im,0}-y_{ir}) \eta_\text{RS} \nonumber\\
  &-\frac{1}{2}\, d_\text{long}\, \varphi_\text{RS} (\varphi_\text{RS}-\varphi_r) \nonumber\\
&\, - \frac{1}{2}\, d_\text{long}\, \eta_\text{RS} (\eta_\text{RS} - \eta_r) \,.
\label{eq:3D_LPSng_pivot_arbitrary}
\end{eqnarray}
Comparing Eq.~\eqref{eq:3D_LPSng_pivot_arbitrary} and Eq.~\eqref{eq:non_geom_3D_beam_const_ximyim}
we see that a rotation around a pivot, that is shifted longitudinally by $d_\text{long}$ against the incidence point ($x_{im},y_{im}$) of the measurement beam on the detector, results in additional first- and second-order non-geometric TTL coupling in $\varphi_\text{RS}$ compared to the case with static offsets. 
The lateral displacement of the centre of rotation $d_\text{lat}$ affects the beam walk only as a secondary effect (Eq.~\eqref{eq: x_im beam CoR}). 
Therefore, %
it does not notably affect the non-geometric TTL coupling for small angles.

\ \\
\begin{tabular}{|ll|}
\br
jitter type: & $\varphi_\text{RS}$ \\
static parameters :& $d_\text{long}$, $d_\text{lat}$ \\
 & $z_{Rm}\, z_{Rr}$, $z_m\, z_r$, $x_{im,0},\,x_{ir}$ \\
detector: &  SEPD \\
\br
\end{tabular}

The description of non-geometric TTL coupling is more complex for unequal beam parameters.
Like above, we substitute $x_{im}$ by its dynamic representation (Eq.~\eqref{eq: x_im beam CoR}) and replace $\varphi_m\rightarrow-\varphi_\text{RS}$ in Eq.~\eqref{eq: non-geom unequal dynamic}. 
Further, we assume for simplicity a nominally impinging reference beam ($\varphi_r=0$) and no detector tilt ($\varphi_\text{PD}=0$), yielding
\begin{eqnarray}
\text{LPS}^\text{SEPD,2D}_\text{ng,RS} %
  &\approx
   -(x_{im,0} - x_{ir}) \frac{z_{Rm} (z_{Rm}+z_{Rr})+z_{m} (z_{m}-z_{r})}{(z_{Rm}+z_{Rr})^2+(z_{m}-z_{r})^2}\, \varphi_\text{RS} \nonumber\\
  &+ d_\text{long}\,(x_{im,0}-x_{ir})\, \left[\frac{(z_m-z_r)}{(z_{Rm}+z_{Rr})^2+(z_m-z_r)^2} \right]\, \varphi_\text{RS} \nonumber\\
   &- d_\text{long} \left[ \frac{z_{Rm} (z_{Rm}+z_{Rr})+z_{m} (z_{m}-z_{r})}{(z_{Rm}+z_{Rr})^2+(z_{m}-z_{r})^2} \right] \, \varphi_\text{RS}^2 \nonumber\\
   &+  (d_\text{long}^2 - d_\text{lat}\,(x_{im,0}-x_{ir})) \left[\frac{(z_m-z_r)}{2\left( (z_{Rm}+z_{Rr})^2+(z_m-z_r)^2\right)} \right] \varphi_\text{RS}^2 
   \nonumber\\
   &- \left[  \frac{(z_{Rr}^2+z_r^2) z_{m}-(z_{Rm}^2+z_m^2) z_r}{(z_{Rm}+z_{Rr})^2+(z_{m}-z_{r})^2}\right] \,\frac{\varphi_\text{RS}^2}{2} \,.
  \label{eq: non-geom unequal SR}
\end{eqnarray}
Hence, the lateral displacement $d_\text{lat}$ couples for unequal beam parameters to a small degree into the readout.

\subsubsection{Laterally jittering receiving system}
\label{sec: non_geom offset equal beam RS lateral}
\ \\
\begin{tabular}{|ll|}
\br
jitter type: & $x_\text{RS}$, $y_\text{RS}$ \\
system parameters: & $\varphi_\text{RS}$, $\eta_\text{RS}$, $d_\text{long}$, $d_\text{lat}$, $d_\text{vert}$ \\
 & $\varphi_r$, $\eta_r$, $\varphi_\text{PD}$, $z_{Rm}=z_{Rr}$, $z_m=z_r$ \\
detector: &  SEPD \\
\br
\end{tabular}

Next, we consider the case of a laterally jittering receiver which is assumed to be angularly misaligned with respect to the incoming beam.  
As explained above, the lateral jitter changes the measurement beam's point of incidence at the detector. 
Since we consider here a receiving system jitter (i.e.\ not measurement beam jitter), it relates to the lateral jitter coupling introduced in Sec.~\ref{sec: non_geom offset equal beam lateral} via $x_m\rightarrow x_\text{RS}$ and $y_m\rightarrow y_\text{RS}$ (compare coordinate system in Figs.~\ref{fig:PD_coordsyst} and \ref{fig:ngeoSCTTL} and the effect of a positive receiver shift on the offset $x_{im}$).

Thus, we find the non-geometric TTL coupling signal by applying the angle and jitter parameter transformations for the receiving system to the Eqs.~\eqref{eq:non-geom_2D_equal_lateral} and \eqref{eq:non_geom_3D_equal_lateral} (setting $\varphi_\text{PD}=0$):
\begin{eqnarray}
\text{LPS}^\text{SEPD,2D}_\text{ng,RS} %
  &\approx  \frac{1}{2}\,x_\text{RS} \left[-(\varphi_\text{RS}+\varphi_\text{PD})+(\varphi_r-\varphi_{\text{PD}})\right]  \,,
\label{eq: non-geom equal lateral PD} \\
\text{LPS}^\text{SEPD,3D}_\text{ng} 
  &\approx  \frac{1}{2}\,x_\text{RS} (-\varphi_\text{RS}+\varphi_r) %
  -      \frac{1}{2}\,y_\text{RS} (\eta_\text{RS}+\eta_r) \,.
\label{eq: non-geom equal lateral 3D}
\end{eqnarray}
Eqs.~\eqref{eq: non-geom equal lateral PD} and~\eqref{eq: non-geom equal lateral 3D} yield significant linear TTL coupling in the lateral jitter parameters depending on the alignment of the interfering beams with respect to the detector surface.

\subsubsection{Full LPS signal as sum of geometric and non-geometric effects}
\label{sec:LPS_cancellation_RS}
\ \\
\begin{tabular}{|ll|}
\br
jitter type: & ($\varphi_\text{RS}$ or $x_\text{RS}$) \\
static parameters: & (- or $\varphi_\text{RS}$), $d_\text{long}$ \\
 &  %
$z_{Rm}=z_{Rr}$, $z_m=z_r$, $x_{im,0}=x_{ir}$ \\ 
detector: &  SEPD \\
\br
\end{tabular}

We now use the equations above to reanalyse a setup previously described in \cite[cf Fig.~9]{G21}.
There, a system rotation with an arbitrary longitudinal offset of the pivot point, but no lateral offset was assumed. Both beams were in the nominal case aligned to each other, i.e.\ $x_{im,0}=x_{ir}$ and $\varphi_r=0$, and we assumed no photodiode tilt, i.e.\ $\varphi_\text{PD}=0$.
Inserting this into Eq.~\eqref{eq: non-geom equal offset beam CoR}, we get
\begin{eqnarray}
\text{LPS}^\text{SEPD,2D}_\text{ng,RS} \approx& -\frac{1}{2} d_\text{long} \varphi_\text{RS}^2 \,.
\label{eq: non geom long SC}
\end{eqnarray}
By comparing Eq.~\eqref{eq: non geom long SC} with the corresponding OPD \cite[cf Eq.~(43)]{G21}, we find that both are equal for a rotation of the system around a longitudinally displaced pivot but have an inverted sign.
Thus, the angular jitter coupling for a longitudinally displaced centre of rotation cancels:
\begin{eqnarray}
\text{LPS}^\text{SEPD,2D}_\text{RS} &=\ \text{OPD}_\text{RS}^\text{2D} + \text{LPS}^\text{SEPD,2D}_\text{ng,RS} \\
&\approx 0\;.%
\label{eq:LPS_SC}
\end{eqnarray}
This confirms the observations found in numerical simulations \cite{G21,Schuster2015}.

Considering instead a laterally jittering receiving system with a constant angular misalignment ($\varphi_\text{RS}\neq0$), we find residual TTL coupling in the complete LPS signal.
Let us, for simplicity, assume a nominally aligned reference beam and detector, i.e.\ $\varphi_r=\varphi_\text{PD}=0$. %
Under these assumptions, Eq.~\eqref{eq: non-geom equal lateral PD} reduces to
\begin{eqnarray}
\text{LPS}^\text{SEPD,2D}_\text{ng,RS} %
&\approx -\frac{1}{2}\,x_\text{RS}\,\varphi_\text{RS}  \,.
\label{eq: non-geom equal lateral simple}
\end{eqnarray}
While seeing a significant non-geometric coupling for lateral jitter, there is no geometric correspondence: 
If the receiver jitters parallelly to its detector surface, the length of the received beam does not change \cite[cf Eq.~(38)]{G21}.
Thus, the lateral TTL coupling here is fully described by the non-geometric coupling,
\begin{eqnarray}
\text{LPS}^\text{SEPD,2D}_\text{RS} = \text{LPS}^\text{SEPD,2D}_\text{ng,RS}%
&\approx -\frac{1}{2}\,x_\text{RS}\,\varphi_\text{RS}  \,,
\label{eq: SC equal lateral}
\end{eqnarray}
which yields a strong linear coupling.
\\ \\
\begin{tabular}{|ll|}
\br
jitter type: & $\varphi_\text{RS}$ \\
static parameters: & $d_\text{long}=z_m$ \\
 &  %
$z_{Rm}$, $z_{Rr}$, $z_m=z_r$, $x_{im,0}=x_{ir}$ \\ 
detector: &  SEPD \\
\br
\end{tabular}

The equations above only hold for the case of equal beam parameters.
If the beam parameters for both interfering beams divert from each other, the cancellation fails (compare Eq.~\eqref{eq: non-geom unequal SR}). 
However, we can construct a special case in which we can relax the conditions on the beam parameters and still achieve a cancellation:
The waist size becomes irrelevant when considering a rotation around the waist position. Let both beams be identical besides their waist size and imping nominally at the non-tilted detector, i.e.\ $z_m=z_r$, $x_{im,0}=x_{ir}$, $\varphi_r=0$ and $\varphi_\text{PD}=0$. The centre of rotation is longitudinally displaced from the detector surface and lies in the beam waist, i.e.\ $d_\text{long}=z_m$. In this case, the geometric and the non-geometric TTL effects cancel each other. 
\begin{eqnarray}
\text{LPS}^\text{SEPD,2D}_\text{RS} 
  &= \text{OPD}^\text{2D}_\text{RS} +\text{LPS}^\text{SEPD,2D}_\text{ng,RS} \\
  &\approx
  \left[ d_\text{long}\,\frac{\varphi_\text{RS}^2}{2} \right]
   - \left[ d_\text{long}\,\varphi_\text{RS} \right] \left[\frac{z_{Rm}}{z_{Rm}+z_{Rr}}\right] \varphi_\text{RS}
   \nonumber\\
  &\, - \left[  \frac{(z_{Rr}-z_{Rm})\, d_\text{long}}{z_{Rm}+z_{Rr}}\right] \,\frac{\varphi_\text{RS}^2}{2}  \\
  &= 0  \,.
  \label{eq: non-geom unequal z0m}
\end{eqnarray}

\subsubsection{Cancellation of the photodiode angle dependent terms}
\label{sec:pd_cancellation_RS}
\ \\
\begin{tabular}{|ll|}
\br
jitter type: & ($\varphi_\text{RS}$ or $x_\text{RS}$) \\
static parameters: & (- or $\varphi_\text{RS}$), $d_\text{long}$, $d_\text{lat}$ \\
 &  $\varphi_\text{PD}$, $z_{Rm}$, $z_{Rr}$, $z_m,\,z_r$ \\ 
detector: &  SEPD \\
\br
\end{tabular}

We have claimed in Sec.~\ref{sec: photodiode angle} that the terms in the full TTL coupling signal depending on the alignment of the detector $\varphi_\text{PD}$ would cancel.
Namely, the $\varphi_\text{PD}$-dependent terms in the non-geometric signal, which all depend on the system jitter induced beam walk, cancel with their geometric counterparts presented in \cite{G21}.
The latter describe the path length change of the measurement beam axis due to the receiver jitter. 
Since the measurement beam walks along the detector surface, a tilt of that surface would shorten or elongate the beam path depending on the sign of the beam tilt \cite{G21}.

We can show this cancellation mathematically for angular jitter coupling by substituting OPD$_\text{RST}^\text{2D}$ (\cite[cf Eq.~(31)]{G21}) and
$x_{im,\text{RS}}^\text{2D}$ (Eq.~\eqref{eq: x_im beam CoR}) into Eq.~\eqref{eq: phi_pd term}.
The same holds for lateral jitter coupling.
Here, we find for the non-geometric coupling contribution
\begin{eqnarray}
\text{LPS}_\text{ng,RS}^\text{SEPD,2D}(\varphi_\text{PD})
-\text{LPS}_\text{ng,RS}^\text{SEPD,2D}(\varphi_\text{PD}=0)
= -x_\text{RS}\,\varphi_\text{PD} \,.
\end{eqnarray}
Adding this to the geometric equation \cite[cf Eq.~(21)]{G21} (there, we need to substitute $y_\text{RS}\rightarrow x_\text{RS}$ due to a different convention), the full signal cancels as well.

\subsection{Mirror jitter}
\label{sec:BeamWalkMechanisms_Mirror}

\begin{figure}
	\centering
	\includegraphics[width= 0.45\columnwidth]{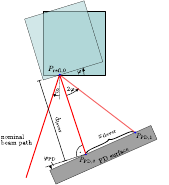} \hspace{3ex}
	\includegraphics[width= 0.45\columnwidth]{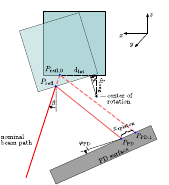}
	\caption{
 Beam walk induced by two TTL mechanisms in the case of mirror rotations.
	Left figure: The lever arm mechanism. Shown is the displacement $x_{i\text{lever}}$ of the point of incidence from $P_\text{PD,0}$ to $P_\text{PD,1}$ due to a tilt $\varphi$ of the mirror around the reflection point $P_\text{refl,0}$. The displacement scales with the distance $d_\text{lever}$ of the photodiode (PD) from the mirror, which is defined along the nominal beam axis (case $\varphi=0$). 
	Right figure: The piston mechanism. The beam's reflection point is translated due to the motion of the mirror surface into the beam path. It is here caused by a shift of the centre of rotation with respect to the reflection point (longitudinally by $d_\text{long}$ and laterally by $d_\text{lat}$, both are positive here). This yields a beam walk $x_{i\text{piston}}$ from point $P_\text{PD,1}$ to point $P_\text{PD}$ additionally to the lever arm beam walk.
	 The dashed line corresponds to the beam path due to the lever arm effect.  
	 In both figures, $\beta$ marks the beam's angle of incidence at the mirror in nominal position and $\varphi_\text{PD}$ denotes the rotation of normal to the PD surface with respect to the nominal beam axis. Arrows pointing clockwise indicate negative angles.}
	\label{fig:ngeoTTL}
\end{figure}

We now assume a mirror that is subject to angular jitter. As one can see on the left-hand side of Fig.~\ref{fig:ngeoTTL}, the path along which the reflected beam propagates to the photodiode is then angle-dependent and deviates from the nominal path, i.e.\ the path for $\varphi = 0$. 
This affects the OPL of the beam axis but also the angle and point of incidence at the detector.
Meanwhile, the second interfering beam is assumed not to be reflected at the mirror but to hold as a static reference.
The measurement beam rotation and beam walk depicted in Fig.~\ref{fig:ngeoTTL} originate from two different mechanisms: the lever arm and the piston mechanism. 
The changes of the beam axis length due to these mechanisms have separately been discussed in \cite{G21}.

The lever arm mechanism (left-hand side of Fig.~\ref{fig:ngeoTTL}) describes the angular jitter of a mirror surface around the beam reflection point $P_\text{refl,0}$. The beam then follows the light red path and hits the detector in an angle-dependent point $P_\text{PD,1}$. 
The distance $x_{i\text{lever}}$ between the original point of incidence $P_\text{PD,0}$ and $P_\text{PD,1}$ describes the lever arm induced beam walk. The change caused by this beam walk in the LPS signal is the non-geometric lever arm effect. 

The piston mechanism (right-hand side of Fig.~\ref{fig:ngeoTTL}) describes all additional 
changes originating from the mirror movement into or out of the beam path, hence changing the beam's point of reflection from $P_\text{refl,0}$ to $P_\text{refl}$.
For angular jitter coupling, this is caused by the  
displacement of the mirror's centre of rotation with respect to the reflection point. 
As a consequence, also the beam's point of incidence at the detector shifts from $P_\text{PD,1}$ to point $P_\text{PD}$. The LPS signal change caused by the beam walk on the photodiode is the non-geometric piston effect.

The piston mechanism also describes lateral jitter coupling. 
Here, the mirror would jitter parallel to the $x$- (or $y$-) axis, also yielding a translation of the point of reflection if the mirror surface is tilted with respect to the axes.

The following subsections show that the equations for the non-geometric lever arm effect are comparable to longitudinal offset in the case of receiver jitter.
However, the piston effect does not only differ from the previous case since we have a reflection (yields approximately twice the OPD) but also since the offset of the centre of rotation from the point of reflection has a much smaller effect on the beam walk.
The same holds for lateral mirror jitter.

\subsubsection{Angularly jittering mirror}
\label{sec: non_geom offset equal beam TM}
\ \\
\begin{tabular}{|ll|}
\br
jitter type: & $\varphi$, $\eta$ \\
static parameters: & $d_\text{lever}$, $d_\text{long}$, $d_\text{lat}$, $d_\text{vert}$, $\beta\equiv\beta_y$, $\beta_z$ \\
 & $\varphi_r$, $\eta_r$, $\varphi_\text{PD}$, $z_{Rm}=z_{Rr}$, $z_m=z_r$, $x_{im,0},\,x_{ir}$, $y_{im,0},\,y_{ir}$ \\
detector: &  SEPD \\
\br
\end{tabular}

We consider the case of a mirror with angular jitter as depicted in Fig.~\ref{fig:ngeoTTL} and an arbitrary pivot location. 
The beam walk $x_{im}$ then depends on the mirror angle $\varphi$, the lever arm $d_\text{lever}$ as well as the longitudinal and lateral displacements between the reflection point and the centre of rotation $d_\text{long}, d_\text{lat}$:
\begin{eqnarray}
	x_{im,\text{MR}}^\text{2D} &\approx \, x_{im,0} 
	-\ 2\left[d_\text{lever}+ d_\text{lat}\sin(\beta)\right] \varphi \nonumber\\
	 &+\  \left[2 d_\text{lat} \cos(\beta)+d_\text{long} \sin(\beta)\right] \varphi^2 \,.
\label{eq: x_im}
\end{eqnarray}
For interfering beams with identical beam parameters we substitute in Eq.~\eqref{eq:non_geom_2D_beam_dynamic} $\varphi_m \rightarrow 2 \varphi$ and replace $x_{im}$ by Eq.~\eqref{eq: x_im} and thus find
\begin{eqnarray}
\text{LPS}^\text{SEPD,2D}_\text{ng,MR} &\approx  
(x_{im,0}-x_{ir}) \,\varphi  \nonumber\\
&- \left[d_\text{lever}+ d_\text{lat}\sin(\beta)\right] \varphi  \,(2\varphi + \varphi_r-2 \varphi_{\text{PD}}) \,.
\label{eq: non-geom equal xim TM}
\end{eqnarray}
We also expand this case for three-dimensional setups. 
The beam's angle of incidence at the mirror is there notated by the propagation angles $\beta_y$ (equivalent to the two-dimensional $\beta$) and $\beta_z$.
For simplicity, we assume a measurement beam with normal incidence ($ \beta_y=\beta_z = 0$) on the mirror and no detector tilt. Then, the linearised tilt-dependent beam walk becomes
\begin{eqnarray}
x_{im,\text{MR}}^\text{3D} &\approx x_{im,0} - 2d_\text{lever}\varphi \,, \\
y_{im,\text{MR}}^\text{3D} &\approx y_{im,0} + 2d_\text{lever}\eta \,.
\label{eq:xim_yim_TM_jitter}
\end{eqnarray}
We can use again Eq.~\eqref{eq:non_geom_3D_beam_dynamic} with $\varphi_m\rightarrow 2\varphi$ and $\eta_m\rightarrow 2\eta$, substitute $x_{im},y_{im}$ by Eq.~\eqref{eq:xim_yim_TM_jitter} and find
\begin{eqnarray}
\text{LPS}^\text{SEPD,3D}_\text{ng,MR} &\approx 
 (x_{im,0}-x_{ir})\,\varphi
-(y_{im,0}-y_{ir})\,\eta \nonumber\\
&- d_\text{lever} \left(2 \varphi^2  + \varphi \, \varphi_r  +2 \eta^2 + \eta \, \eta_r\right)  
\label{eq: non-geom equal xim TM_3D}
\end{eqnarray}

As shown by Eqs.~\eqref{eq: non-geom equal xim TM} and~\eqref{eq: non-geom equal xim TM_3D}, the mirror jitter couples to first- as well as second-order into the readout.
For any given lever arm $d_\text{lever}$ or a lateral offset $d_\text{lat}$, an additional linear effect originates from an angular misalignment of the reference beam or a photodiode tilt. If both are optimally aligned, the beam walk would only yield second-order non-geometric TTL coupling in $\varphi$ (and $\eta$).

\ \\
\begin{tabular}{|ll|}
\br
jitter type: & $\varphi$ \\
system parameters: & $d_\text{lever}$, $d_\text{long}$, $d_\text{lat}$, $\beta$ \\
 & $z_{Rm},\,z_{Rr}$, $z_m,\,z_r$, $x_{im,0},\,x_{ir}$ \\
detector: &  SEPD \\
\br
\end{tabular}

Analogously we proceed 
for the case of beams with unequal parameters.
Here, we substitute in Eq.~\eqref{eq: non-geom unequal dynamic} $x_{im}$ by Eq.~\eqref{eq: x_im} and $\varphi_m\rightarrow2\varphi$. 
Additionally, we set $\varphi_r=\varphi_\text{PD}=0$ for simplicity.
Thus we find
\begin{eqnarray}
& \text{LPS}^\text{SEPD,2D}_\text{ng,MR} \nonumber\\
  &\approx
   \ 2 (x_{im,0} - x_{ir}) \frac{z_{Rm} (z_{Rm}+z_{Rr})+z_{m} (z_{m}-z_{r})}{(z_{Rm}+z_{Rr})^2+(z_{m}-z_{r})^2}\, \varphi \nonumber\\
   &- \, 2\left[d_\text{lever}+ d_\text{lat}\sin(\beta)\right]\,(x_{im,0}-x_{ir})\, \left[\frac{(z_m-z_r)}{(z_{Rm}+z_{Rr})^2+(z_m-z_r)^2} \right] \varphi \nonumber\\
   &-\, 4\left[d_\text{lever}+ d_\text{lat}\sin(\beta)\right] \left[ \frac{z_{Rm} (z_{Rm}+z_{Rr})+z_{m} (z_{m}-z_{r})}{(z_{Rm}+z_{Rr})^2+(z_{m}-z_{r})^2} \right] \varphi^2 \nonumber\\
   &+ \, \left\{2\left[d_\text{lever}+ d_\text{lat}\sin(\beta)\right]^2 + \left[2 d_\text{lat} \cos(\beta)+d_\text{long} \sin(\beta)\right]\,(x_{im,0}-x_{ir})\right\}\, \nonumber\\
   &\quad \cdot\left[\frac{(z_m-z_r)}{(z_{Rm}+z_{Rr})^2+(z_m-z_r)^2} \right] \varphi^2 
   \nonumber\\
   &- 2 \left[  \frac{(z_{Rr}^2+z_r^2) z_{m}-(z_{Rm}^2+z_m^2) z_r}{(z_{Rm}+z_{Rr})^2+(z_{m}-z_{r})^2}\right] \varphi^2 \,.
  \label{eq: non-geom unequal MR} 
\end{eqnarray}

\subsubsection{Laterally jittering mirror}
\label{sec: non_geom offset equal beam TM lateral}
\ \\
\begin{tabular}{|ll|}
\br
jitter type: & $d_\text{lat}(t)$, $d_\text{vert}(t)$ \\
static parameters: & $\varphi$, $\eta$, $\beta\equiv\beta_y,\,\beta_z$ \\
 & $\varphi_r$, $\eta_r$, $\varphi_\text{PD}$, $z_{Rm},\,z_{Rr}$, $z_m,\, z_r$, $x_{im,0},\,x_{ir}$, $y_{im,0},\,y_{ir}$ \\
detector: &  SEPD \\
\br
\end{tabular}

The equations \eqref{eq: x_im}-\eqref{eq: non-geom unequal MR}
hold for angular jitter coupling assuming time-dependent rotations $\varphi(t)$ and $\eta(t)$ as well as for lateral jitter coupling assuming a time-dependency of $d_\text{lat}(t)$ and $d_\text{vert}(t)$. 
We see in Eq.~\eqref{eq: non-geom equal xim TM_3D} that lateral jitter would not couple into the non-geometric LPS signal if the measurement beam has a normal incidence on the mirror, i.e.\ $\beta_y,\beta_z=0$. 
Contrary, Eq.~\eqref{eq: non-geom equal xim TM} shows that the lateral behaviour enters the signal if $\beta\neq 0$. 
However, this coupling is small for small lateral jitter 
yielding in total a third-order effect.
We thus find
\begin{eqnarray}
\text{LPS}^\text{SEPD,2D}_\text{ng,MR} &\approx 0 \,.
\label{eq: non-geom equal xim TM lateral}
\end{eqnarray}

\subsubsection{Full LPS Signal as sum of geometric and non-geometric effects}
\label{sec:LPS_cancellation_MR}
\ \\
\begin{tabular}{|ll|}
\br
jitter type: & ($\varphi$ or $d_\text{lat}(t)$) \\
static parameters: & ($d_\text{lever}$, $d_\text{long}$, $d_\text{lat}$ or $\varphi$), $\beta$ \\
 &  %
$z_{Rm}=z_{Rr}$, $z_m=z_r$, $x_{im,0}=x_{ir}$ \\ 
detector: &  SEPD \\
\br
\end{tabular}

Now, we can combine the non-geometric and the geometric TTL contributions to see the total effect and do this here for simplicity only for the two-dimensional case.
Investigating the case of ideal alignment, i.e.\ $x_{im,0}=x_{ir}$, $\varphi_r=0$, and using Eq.~(10) from \cite{G21}, we find a total LPS of
\begin{eqnarray}
\text{LPS}^\text{SEPD,2D}_\text{MR} &=\ \text{OPD}_\text{MR}^\text{2D} + \text{LPS}^\text{SEPD,2D}_\text{ng,MR} \\
&\approx -2 d_\text{lat} \cos(\beta) \varphi 
	+ d_\text{long} \cos(\beta) \varphi^2 \,.
\label{eq:LPS_MR_cancellation}
\end{eqnarray}
Thus, the geometric and non-geometric lever arm effects cancel in the total TTL equations.
However, the non-geometric piston effect does not fully cancel its geometric counterpart leaving the residual given in Eq.~\eqref{eq:LPS_MR_cancellation}.
The same holds for the more general case of arbitrarily aligned beams.
In conclusion, we see that 
the total piston effect will be the dominant noise source since the total lever arm effect is zero. 
A lateral displacement $d_\text{lat}$ of the centre of rotation relative to the incidence point on the mirror causes linear coupling and should be avoided if possible. Contrary, a longitudinal displacement of the centre of rotation adds only second-order coupling and is, therefore, less critical.

Interpreting the equation for lateral jitter coupling (compare Sec.~\ref{sec: non_geom offset equal beam TM lateral}), we see by Eq.~\eqref{eq:LPS_MR_cancellation} that the uncancelled geometric lateral jitter coupling $d_\text{lat}(t)$ induces significant TTL coupling.

\subsubsection{Cancellation of the photodiode angle dependent terms}
\label{sec:pd_cancellation_MR}
\ \\
\begin{tabular}{|ll|}
\br
jitter type: & ($\varphi$ or $d_\text{lat}(t)$) \\
static parameters: & ($d_\text{lever}$, $d_\text{long}$ or $d_\text{lat}$ or $\varphi$), $\beta$ \\
 & $\varphi_\text{PD}$, $z_{Rm},\,z_{Rr}$, $z_m,\,z_r$, $x_{im,0},\,x_{ir}$ \\
detector: &  SEPD \\
\br
\end{tabular}

Analogously to the case of receiver jitter, we can prove the cancellation of the $\varphi_\text{PD}$-dependent terms simply by substitution of the presented equations into Eq.~\eqref{eq: phi_pd term}.
For angular mirror jitter, this means replacing OPD$_\text{MRT}^\text{2D}$ by \cite[cf Eq.~(30)]{G21} and $x_{im}$ by the representation $x_{im,\text{MR}}^\text{2D}$ (Eq.~\eqref{eq: x_im}).
For lateral jitter coupling, we find a priori neither for geometric nor for non-geometric coupling non-negligible detector tilt dependent terms.

\subsection{Propagation of tilted beams trough transmissive components}
In addition to the two systems introduced above, transmissive components can be deployed along the measurement beam path.
The beam refraction occurring due to the different refractive indices of the surrounding medium and the component's material changes the beam path within the component. 
Consequently, the beam is shifted laterally with respect to the path the beam would propagate if the transmissive component was not there.

This shift changes with the alignment of the beam before the transmission and hence with angular jitter.
This generates a beam walk and, by this, non-geometric TTL coupling.
In this section, we discuss this coupling analytically.

\subsubsection{Transmissive components along the path of the angular jittering beam}
\label{sec: non_geom offset equal beam tc}
\ \\
\begin{tabular}{|ll|}
\br
jitter type: & $\varphi_m$ \\
static parameters: & $t_\text{BS}$, $n_\text{BS}$, $\varphi_\text{BS}$ \\
 & $\varphi_r$, $\varphi_\text{PD}$, $z_{Rm},\,z_{Rr}$, $z_m,\,z_r$, $x_{im,0},\,x_{ir}$ \\
detector: &  SEPD \\
\br
\end{tabular}

The effect of transmissive components on the measurement beam path has previously been discussed in \cite{G21}.
Analogously to that case, we assume here a transmission of the measurement beam through $n$ components with thicknesses $t_{\text{BS},i}$, refraction indices $n_{\text{BS},i}$ and rotation of $\varphi_{\text{BS},i}$ of the surface normal with respect to the nominal incoming beam path.
The resulting beam walk is
\begin{eqnarray}
x_{im,\text{tc}}^\text{2D} &\approx 
 - \sum_i t_{\text{BS},i} \left[ \frac{n_{\text{BS},i}^2 \cos(\varphi_{\text{BS},i})^2}{(n_{\text{BS},i}^2-\sin^2(\varphi_{\text{BS},i}))^{3/2}} - \frac{\sin^2(\varphi_{\text{BS},i})}{\sqrt{n_{\text{BS},i}^2-\sin^2(\varphi_{\text{BS},i})}} -\cos(\varphi_{\text{BS},i}) \right] \varphi_m \nonumber\\
 &-\  \sum_i \frac{3}{2}\, \frac{n_{\text{BS},i}^2 t_{\text{BS},i} \cos(\varphi_{\text{BS},i}) \sin(\varphi_{\text{BS},i})}{((n_{\text{BS},i}^2-\sin^2(\varphi_{\text{BS},i}))^{5/2}} \left(n_{\text{BS},i}^2-1\right) \, \varphi_m^2 \ .
\label{eq: x_im tc}
\end{eqnarray}
This equation can be used both for the case of angular jitter of a receiver (when the components jitter simultaneously to the receiver, replace $\varphi_m\rightarrow-\varphi_\text{RS}$) or of a mirror (for components transmitted after the reflection $\varphi_m\rightarrow2\varphi$). 
The total beam walk is then the sum of Eq.~\eqref{eq: x_im tc} and the offsets derived for a rotation of the setup (Eq.~\eqref{eq: x_im beam CoR}) or the mirror (Eq.~\eqref{eq: x_im}):
\begin{eqnarray}
x_{im,\text{RST}} &= x_{im,\text{RS}} + x_{im,\text{tc}} \,.
\label{eq:xim_RST}\\
x_{im,\text{MRT}} &= x_{im,\text{MR}} + x_{im,\text{tc}} \,, 
\label{eq:xim_MRT}
\end{eqnarray}
In the given case of identical beam parameters, this sum of beam walks also results in a full non-geometric LPS signal which is a sum of the previously derived terms, and an additional LPS signal due to the transmissive component:
\begin{eqnarray}
\text{LPS}_{\text{ng,RST}} &= \text{LPS}_{\text{ng,RS}} + \text{LPS}^\text{SEPD}_\text{ng,tc}  \,,  \\
\text{LPS}_{\text{ng,MRT}} &= \text{LPS}_{\text{ng,MR}} + \text{LPS}^\text{SEPD}_\text{ng,tc}  \,.
\label{eq:LPS_MRT}
\end{eqnarray}
In the two-dimensional case, the additive coupling term reads for equal beam parameters
\begin{eqnarray}
\text{LPS}^\text{SEPD,2D}_\text{ng,tc}
&\approx -\sum_i \frac{t_{\text{BS},i}}{2} &\Biggg[ \frac{n_{\text{BS},i}^2 \cos(\varphi_{\text{BS},i})^2}{(n_{\text{BS},i}^2-\sin^2(\varphi_{\text{BS},i}))^{3/2}} - \frac{\sin^2(\varphi_{\text{BS},i})}{\sqrt{n_{\text{BS},i}^2-\sin^2(\varphi_{\text{BS},i})}} \nonumber\\
& & -\cos(\varphi_{\text{BS},i}) \Biggg] 
\varphi_m \left[\varphi_m+\varphi_r-2\varphi_{\text{PD}}\right] 
\label{eq:LPSng_tc}
\end{eqnarray} 
and otherwise
\begin{eqnarray}
& \text{LPS}^\text{SEPD,2D}_\text{ng,tc} \nonumber\\
  \approx&
   \ (x_{im,0} - x_{ir}) \frac{z_{Rm} (z_{Rm}+z_{Rr})+z_{m} (z_{m}-z_{r})}{(z_{Rm}+z_{Rr})^2+(z_{m}-z_{r})^2} \varphi_m \nonumber\\
  -& \, \sum_i t_{\text{BS},i} \left[ \frac{n_{\text{BS},i}^2 \cos(\varphi_{\text{BS},i})^2}{(n_{\text{BS},i}^2-\sin^2(\varphi_{\text{BS},i}))^{3/2}} - \frac{\sin^2(\varphi_{\text{BS},i})}{\sqrt{n_{\text{BS},i}^2-\sin^2(\varphi_{\text{BS},i})}} -\cos(\varphi_{\text{BS},i}) \right] \nonumber\\
  &\qquad \cdot(x_{im,0}-x_{ir})\, \left[\frac{(z_m-z_r)}{(z_{Rm}+z_{Rr})^2+(z_m-z_r)^2} \right] \varphi_m \nonumber\\
   -&\, \sum_i t_{\text{BS},i} \left[ \frac{n_{\text{BS},i}^2 \cos(\varphi_{\text{BS},i})^2}{(n_{\text{BS},i}^2-\sin^2(\varphi_{\text{BS},i}))^{3/2}} - \frac{\sin^2(\varphi_{\text{BS},i})}{\sqrt{n_{\text{BS},i}^2-\sin^2(\varphi_{\text{BS},i})}} -\cos(\varphi_{\text{BS},i}) \right] \nonumber\\
   &\qquad \cdot\left[ \frac{z_{Rm} (z_{Rm}+z_{Rr})+z_{m} (z_{m}-z_{r})}{(z_{Rm}+z_{Rr})^2+(z_{m}-z_{r})^2} \right] \varphi_m^2 \nonumber\\
  +&  \Biggg\{ \left( \sum_i t_{\text{BS},i} \left[ \frac{n_{\text{BS},i}^2 \cos(\varphi_{\text{BS},i})^2}{(n_{\text{BS},i}^2-\sin^2(\varphi_{\text{BS},i}))^{3/2}} - \frac{\sin^2(\varphi_{\text{BS},i})}{\sqrt{n_{\text{BS},i}^2-\sin^2(\varphi_{\text{BS},i})}} -\cos(\varphi_{\text{BS},i}) \right]\right)^2 \nonumber\\
 &\qquad - \sum_i \frac{3}{2}\, \frac{n_{\text{BS},i}^2 t_{\text{BS},i} \cos(\varphi_{\text{BS},i}) \sin(\varphi_{\text{BS},i})}{((n_{\text{BS},i}^2-\sin^2(\varphi_{\text{BS},i}))^{5/2}} \left(n_{\text{BS},i}^2-1\right) \,(x_{im,0}-x_{ir}) \Biggg\} \nonumber\\
  &\qquad \cdot\left[\frac{(z_m-z_r)}{(z_{Rm}+z_{Rr})^2+(z_m-z_r)^2} \right] \varphi_m^2 
   \nonumber\\
  -& \left[  \frac{(z_{Rr}^2+z_r^2) z_{m}-(z_{Rm}^2+z_m^2) z_r}{(z_{Rm}+z_{Rr})^2+(z_{m}-z_{r})^2}\right] \frac{\varphi_m^2}{2} \,.
  \label{eq: non-geom unequal tc} 
\end{eqnarray}

\subsubsection{LPS signal as sum of geometric and non-geometric effects}
\label{sec:LPS_cancellation_tc}
\ \\
\begin{tabular}{|ll|}
\br
jitter type: & $\varphi_m$ \\
system parameters: & $t_\text{BS}$, $n_\text{BS}$, $\varphi_\text{BS}$ \\
static parameters: &  %
$z_{Rm}=z_{Rr}$, $z_m=z_r$, $x_{im,0}=x_{ir}$ \\ 
detector: &  SEPD \\
\br
\end{tabular}

For equal beam parameters, transmissive components equally affect the geometric TTL coupling \cite[cf Eq.~(26)]{G21}. 
It can then be shown that the transmissive component dependent geometric and non-geometric TTL terms cancel in the full TTL signal
\begin{eqnarray}
\text{LPS}^\text{SEPD,2D}_\text{tc} &=\, \text{OPD}_\text{tc}^\text{2D} + \text{LPS}^\text{SEPD,2D}_\text{ng,tc}  \\
&\approx\, 0 \,,
\label{eq:LPS_tc}
\end{eqnarray}
provided that the reference beam is nominally aligned, i.e.\ $ \varphi_r = 0$.

In summary, transmissive components do not contribute to angular jitter TTL coupling if both beams feature the same beam parameters and the reference beam impinges with a normal incidence at the detector.
However, for unequal beam parameters, this cancellation is imperfect.

\subsubsection{Cancellation of the detector angle dependent terms}
\label{sec:pd_cancellation_tc}
\ \\
\begin{tabular}{|ll|}
\br
jitter type: & $\varphi_m$ \\
static parameters: & $t_\text{BS}$, $n_\text{BS}$, $\varphi_\text{BS}$ \\
 & $\varphi_\text{PD}$, $z_{Rm},\,z_{Rr}$, $z_m,\,z_r$, $x_{im,0},\,x_{ir}$ \\
detector: &  SEPD \\
\br
\end{tabular}

The cancellation of the detector tilt dependent terms also holds if transmissive components are placed along the measurement beam path.
Therefore, we compare in the case of a mirror rotation substituting OPD$_\text{MRT}^\text{2D}$ (\cite[cf Eq.~(30)]{G21}) for the geometric and LPS$_\text{ng}^\text{SEPD,2D}$ (Eq.~\eqref{eq: non-geom unequal}) for the non-geometric contribution with $x_{im,\text{MRT}}^\text{2D}$ (Eq.~\eqref{eq:xim_MRT}). 
In the case of a rotation of the receiving system we substitute OPD$_\text{RST}^\text{2D}$ (\cite[cf Eq.~(31)]{G21})  and LPS$_\text{ng}^\text{SEPD,2D}$ (Eq.~\eqref{eq: non-geom unequal}) with $x_{im,\text{RST}}^\text{2D}$ (Eq.~\eqref{eq:xim_RST}).

\section{Summary non-geometric TTL coupling}
\label{sec:summary}
In this section, we summarise the non-geometric TTL effects providing an estimate of the polynomial degree of the added TTL coupling. 
Throughout this paper, we have mostly assumed single element photodiodes, and therefore also assume this detector type here. However, we have shown in \cref{sec: detector size} that these results are relevant also when quadrant photodiodes are used, and that the deviation for QPDs and the LPS$^\text{LPF}$ signal from the given equations is particularly small.
For simplicity, we assume here a nominally aligned measurement and reference beam, i.e.\ $\beta=0,\,\varphi_r=0$. Furthermore, we assume no detector tilt, i.e.\ $\varphi_\text{PD}=0$, since we have shown in \cref{sec: photodiode angle} that the detector tilt angle cancels in either case from the total LPS signal.

\subsection{Angular jitter coupling}
Given two beams that are nominally aligned and always share the same point of incidence, the non-geometric coupling signal for equal beam parameters becomes negligible. However, we face TTL coupling if both beams feature different beam parameters (compare Eq.~\eqref{eq: non-geom unequal}), i.e.
\begin{eqnarray}
  \text{LPS}^\text{SEPD,2D}_\text{ng}  =&
  - \,
  \left[  \frac{(z_{Rr}^2+z_r^2) z_{m}-(z_{Rm}^2+z_m^2) z_r}{(z_{Rm}+z_{Rr})^2+(z_{m}-z_{r})^2}\right] \frac{\varphi_m^2}{2}
 \, .
 \label{eq:sum_non-geom_unequal}
\end{eqnarray}

For equal and for unequal beam parameters, a static offset between the measurement and the reference beam ($x_{im}-x_{ir}=\text{const.}$) makes the jitter couple linearly into the signal. This holds for the case of a rotating receiver as well as for a rotating mirror (compare Eq.~\eqref{eq: non-geom equal offset}). Under our assumptions, we get for equal beam parameters
\begin{eqnarray}
\text{LPS}^\text{SEPD,2D}_\text{ng} \approx&\, (x_{im}-x_{ir}) \frac{\varphi_m}{2} \,
\label{eq:sum_non-geom_offset}
\end{eqnarray} 
and for unequal beam parameters (compare Eq.~\eqref{eq: non-geom unequal})
\begin{eqnarray}
& \text{LPS}^\text{SEPD,2D}_\text{ng} \nonumber\\ &\approx
   \ (x_{im} - x_{ir}) \left[ \frac{z_{Rm} (z_{Rm}+z_{Rr})+z_{m} (z_{m}-z_{r})}{(z_{Rm}+z_{Rr})^2+(z_{m}-z_{r})^2} \right] \varphi_m
   \nonumber\\
    &- \,
  \left[  \frac{(z_{Rr}^2+z_r^2) z_{m}-(z_{Rm}^2+z_m^2) z_r}{(z_{Rm}+z_{Rr})^2+(z_{m}-z_{r})^2}\right] \frac{\varphi_m^2}{2} \,.
  \label{eq:sum_non-geom_offset_unequal}
\end{eqnarray}
Furthermore, the TTL coupling in the case of a setup dependent beam walk, i.e.\ if the centre of rotation is not located on the photodiode surface, adds second-order TTL coupling both for the case of angular jitter of the receiving system (RS, cf. Eq.~\eqref{eq: non-geom equal offset beam CoR}) or a reflecting mirror (MR, cf. Eq.~\eqref{eq: non-geom equal xim TM}). 
We find for beams with equal beam parameters:
\begin{eqnarray}
	&\text{LPS}^\text{SEPD,2D}_\text{ng,RS}(\varphi_\text{RS}(t)) &\approx 
	-\frac{1}{2}\, d_\text{long}\, \varphi_\text{RS}^2 \,,
	\label{eq:sum_non-geom_dynamic_off_SC} \\
	&\text{LPS}^\text{SEPD,2D}_\text{ng,MR}(\varphi(t)) &\approx 
	\, -2d_\text{lever} \varphi^2 \,.
	\label{eq:sum_non-geom_dynamic_off_TM}
\end{eqnarray}
Transmissive components also contribute as second-order TTL coupling, no matter whether angular jitter of a mirror or a receiving system is considered. For equal beam parameters we find (cf. Eq.~\eqref{eq:LPSng_tc}):
\begin{eqnarray}
\text{LPS}^\text{SEPD,2D}_\text{ng,tc} \approx&  
\sum_i t_{\text{BS},i} \Biggg[ \frac{n_{\text{BS},i}^2 \cos(\varphi_{\text{BS},i})^2}{(n_{\text{BS},i}^2-\sin^2(\varphi_{\text{BS},i}))^{3/2}} - \frac{\sin^2(\varphi_{\text{BS},i})}{\sqrt{n_{\text{BS},i}^2-\sin^2(\varphi_{\text{BS},i})}} \nonumber\\
&\quad -\cos(\varphi_{\text{BS},i}) \Biggg] \, \frac{\varphi_m^2}{2} \,.
\label{eq:sum_non-geom_dynamic_off_tc}
\end{eqnarray}
For non-equal beam parameters, we find not only second-order but also linear coupling since we get terms that depend on the static beam offsets and the dynamic beam walk. However, assuming a negligible static beam offsets, Eqs.~\eqref{eq: non-geom unequal SR}, \eqref{eq: non-geom unequal MR} and \eqref{eq: non-geom unequal tc} reduce to second-order coupling equations.
\begin{eqnarray}
& \text{LPS}^\text{SEPD,2D}_\text{ng,RS} (\varphi_\text{RS}(t)) \nonumber\\
  \approx&
   \, -d_\text{long} \left[ \frac{z_{Rm} (z_{Rm}+z_{Rr})+z_{m} (z_{m}-z_{r})}{(z_{Rm}+z_{Rr})^2+(z_{m}-z_{r})^2} \right] \varphi_\text{RS}^2 \nonumber\\
  +&  \, d_\text{long}^2\, \left[\frac{(z_m-z_r)}{2\left( (z_{Rm}+z_{Rr})^2+(z_m-z_r)^2\right)} \right] \varphi_\text{RS}^2 
   \nonumber\\
  -& \left[  \frac{(z_{Rr}^2+z_r^2) z_{m}-(z_{Rm}^2+z_m^2) z_r}{(z_{Rm}+z_{Rr})^2+(z_{m}-z_{r})^2}\right] \frac{\varphi_\text{RS}^2}{2} \,,
  \label{eq:sum_non-geom_unequal_SR}
\end{eqnarray}
\begin{eqnarray}
& \text{LPS}^\text{SEPD,2D}_\text{ng,MR} (\varphi(t)) \nonumber\\
  \approx&
    -4d_\text{lever} \bigg[ \frac{z_{Rm} (z_{Rm}+z_{Rr})+z_{m} (z_{m}-z_{r})}{(z_{Rm}+z_{Rr})^2+(z_{m}-z_{r})^2} \bigg] \varphi^2 \nonumber\\
  +& \, 2d_\text{lever}^2 \, \left[\frac{(z_m-z_r)}{(z_{Rm}+z_{Rr})^2+(z_m-z_r)^2} \right] \varphi^2 
   \nonumber\\
  -& 2 \left[  \frac{(z_{Rr}^2+z_r^2) z_{m}-(z_{Rm}^2+z_m^2) z_r}{(z_{Rm}+z_{Rr})^2+(z_{m}-z_{r})^2}\right] \varphi^2 \,,
  \label{eq:sum_non-geom_unequal_MR} 
\end{eqnarray}
\begin{eqnarray}
& \text{LPS}^\text{SEPD,2D}_\text{ng,tc} \nonumber\\
  \approx&
  - \sum_i t_{\text{BS},i} \left[ \frac{n_{\text{BS},i}^2 \cos(\varphi_{\text{BS},i})^2}{(n_{\text{BS},i}^2-\sin^2(\varphi_{\text{BS},i}))^{3/2}} - \frac{\sin^2(\varphi_{\text{BS},i})}{\sqrt{n_{\text{BS},i}^2-\sin^2(\varphi_{\text{BS},i})}} -\cos(\varphi_{\text{BS},i}) \right] \nonumber\\
  &\ \cdot\left[ \frac{z_{Rm} (z_{Rm}+z_{Rr})+z_{m} (z_{m}-z_{r})}{(z_{Rm}+z_{Rr})^2+(z_{m}-z_{r})^2} \right] \varphi_m^2 \nonumber\\
  +& \left( \sum_i t_{\text{BS},i} \left[ \frac{n_{\text{BS},i}^2 \cos(\varphi_{\text{BS},i})^2}{(n_{\text{BS},i}^2-\sin^2(\varphi_{\text{BS},i}))^{3/2}} - \frac{\sin^2(\varphi_{\text{BS},i})}{\sqrt{n_{\text{BS},i}^2-\sin^2(\varphi_{\text{BS},i})}} -\cos(\varphi_{\text{BS},i}) \right]\right)^2 \nonumber\\
  &\ \cdot\left[\frac{(z_m-z_r)}{(z_{Rm}+z_{Rr})^2+(z_m-z_r)^2} \right] \varphi_m^2 
   \nonumber\\
  -& \left[  \frac{(z_{Rr}^2+z_r^2) z_{m}-(z_{Rm}^2+z_m^2) z_r}{(z_{Rm}+z_{Rr})^2+(z_{m}-z_{r})^2}\right] \frac{\varphi_m^2}{2} \,.
  \label{eq:sum_non-geom_unequal_tc} 
\end{eqnarray}

\subsection{Lateral jitter coupling}
Lateral jitter does not cause any first- or second-order non-geometric TTL coupling in the case of a jittering mirror and under the given assumptions:
\begin{eqnarray}
\text{LPS}^\text{SEPD,2D}_\text{ng,MR}(d_\text{lat}(t))  \approx 0 \,.
\label{eq:sum_non-geom_lateral-jitter}
\end{eqnarray}
However, this is different for a lateral jittering receiving system.
A laterally jittering receiving system changes the point of incidence of the measurement beam ($x_{im}$) with respect to the reference beam. 
Therefore, assuming a constant misalignment of the receiver with respect to the received beam and a varying point of incidence $x_{im}(t)$ of the latter, yields a strong linear TTL coupling for equal beam parameters by Eq.~\eqref{eq: non-geom equal lateral simple}
\begin{eqnarray}
& \text{LPS}^\text{SEPD,2D}_\text{ng,RS}(x_\text{RS}) \approx -\frac{1}{2}x_\text{RS}\, \varphi_\text{RS} \,
\label{eq:sum_non-geom_lateral-jitter_SRequal}
\end{eqnarray}
as well as for unequal beam parameters
\begin{eqnarray}
& \text{LPS}^\text{SEPD,2D}_\text{ng,RS}(x_\text{RS}(t)) \nonumber\\ &\approx
   -x_\text{RS}(t) \left[ \frac{z_{Rm} (z_{Rm}+z_{Rr})+z_{m} (z_{m}-z_{r})}{(z_{Rm}+z_{Rr})^2+(z_{m}-z_{r})^2} \right] \varphi_\text{RS}
 \,.
\label{eq:sum_non-geom_lateral-jitter_SRunequal}
\end{eqnarray}

For the given assumptions, all presented non-geometric TTL effects are summarised in Tab.~\ref{tab: TTL_NG_overview}. This table is the counterpart to the geometric TTL effects summarised in \cite[cf Tab.~1]{G21}.

\fulltable{Overview of the different non-geometric cross-coupling mechanisms for a beam with normal incidence and no photodiode tilt (except where explicitly stated otherwise), i.e.\ $\beta=\varphi_\text{PD}=0$. We further assume normal incidence of the reference beam, i.e.\ $\varphi_r=0$, and equal beam parameters, unless explicitly stated otherwise. For each effect we give a short description and the general behaviour (approximated), like linear, quadratic or mixed with respect to the tilt angle. Cross-coupling due to wavefront errors and detector geometry have in general an arbitrary form and are reported here only for completeness. The non-geometric effects apply both for mirror rotation (MR) or receiving system rotation (RS), unless explicitly mentioned otherwise. 
	}
	\centering
	\label{tab: TTL_NG_overview}
	\renewcommand{\arraystretch}{1.5}
\begin{tabular*}{1\textwidth}{@{\extracolsep{\fill}}>{\raggedright}l| l| >{\raggedright}p{0.155\textwidth} c c c c p{0.35\textwidth}}
	\br
	& & Cross-coupling mechanism  &  Name  &  \parbox[t]{15.5mm}{\centering General behaviour} &  Eq.  &  Sec.  &  Description
	\\
	\hline
	\parbox[t]{2.5mm}{\multirow{17}{*}{\rotatebox[origin=c]{90}{wavefront dependent TTL}}}
	& \parbox[t]{2.5mm}{\multirow{4}{*}{\rotatebox[origin=c]{90}{lateral}}}
	& receiver jiter & LPS$_\text{ng,RS}^\text{SEPD}$ 
	& linear & \eqref{eq:sum_non-geom_lateral-jitter_SRequal},\eqref{eq:sum_non-geom_lateral-jitter_SRunequal} & \ref{sec: non_geom offset equal beam RS lateral} & 
	Lateral jitter yields a beam walk of the measurement beam. \\
	& & mirror jitter & LPS$_\text{ng,MR}^\text{SEPD}$ 
	& negligible & \eqref{eq:sum_non-geom_lateral-jitter} & \ref{sec: non_geom offset equal beam TM lateral} & 
	Lateral jitter coupling affects the non-geometric LPS only at higher orders. \\\cline{2-8}
	& \parbox[t]{2.5mm}{\multirow{9}{*}{\rotatebox[origin=c]{90}{angular}}}
	& beam offset & LPS$_\text{ng}^\text{SEPD}$ & mixed & \eqref{eq:sum_non-geom_offset},\eqref{eq:sum_non-geom_offset_unequal}
	& \ref{sec: non_geom offset equal beam} & Initial misalignment on the detector generates asymmetric disparity. \\ [-3pt]	
	& & \multirow{5}{*}{}beam walk & & & & & \\ [-3pt]
	& & $-$ receiver & LPS$_\text{ng,RS}^\text{SEPD}$  & quadratic 
	& \eqref{eq:sum_non-geom_dynamic_off_SC},\eqref{eq:sum_non-geom_unequal_SR}
	& \ref{sec: SR-arbitrary_pivot_identical_Gaussians} 
	& Offsets between rotation point and detector lead to angle dependent beam walk.  \\ [-3pt]
	& & $-$ mirror & LPS$_\text{ng,MR}^\text{SEPD}$  & quadratic 
	& \eqref{eq:sum_non-geom_dynamic_off_TM},\eqref{eq:sum_non-geom_unequal_MR}
	& \ref{sec: non_geom offset equal beam TM}
	& Same as for RS beam walk.  \\ [-3pt] 
	& & $-$ \parbox[t]{8mm}{transmissive\\ components} & LPS$_\text{ng,tc}^\text{SEPD}$  & quadratic 
	& \eqref{eq:sum_non-geom_dynamic_off_tc},\eqref{eq:sum_non-geom_unequal_tc} 
	& \ref{sec: non_geom offset equal beam tc} 
	& Transmissive optical components lead to an additional angle dependent beam walk.
	\\
	& & beam parameters & LPS$_\text{ng}^\text{SEPD}$  & quadratic & \eqref{eq:sum_non-geom_unequal} & \ref{sec: non_geom beam parameter} & {Tilting wavefronts with a curvature mismatch generates coupling.} 
	\\\cline{2-8}
	& \parbox[t]{2.5mm}{\multirow{4}{*}{\rotatebox[origin=c]{90}{both}}}
	& reference beam tilt &  & linear & & \ref{sec: wavefront geometry} & Reference beam tilts add in all cases linear coupling.  
	\\ 
	& & wavefront errors &  & arbitrary & & \ref{sec: higher oder modes} & Aberrations in the wavefronts disturb the balance between different detector sides.  
	\\ 
	\hline
	\parbox[t]{2.5mm}{\multirow{5}{*}{\rotatebox[origin=c]{90}{detector TTL}}}
	& \parbox[t]{2.5mm}{\multirow{5}{*}{\rotatebox[origin=c]{90}{both}}}
	& detector geometry  &  & arbitrary & & \ref{sec:detector_errors} & Errors and additional detector features alter the measured results.  
	\\
	& & tilt of detector &  & negligible & \eqref{eq: phi_pd term} & \ref{sec: photodiode angle} & Tilting the detector effects the geometric and non-geometric cross-coupling inversely. Hence its contribution to the full signal cancels.
	\\
	\br
\end{tabular*}
\endfulltable

\section{Key findings on total TTL coupling}
\label{sec: overview}
After summarising the non-geometric TTL effects in the previous section, we now discuss our findings on the total TTL effect, i.e.\ the sum of geometric and non-geometric TTL effects. 

We have demonstrated in Secs.~\ref{sec:LPS_cancellation_RS} and \ref{sec:LPS_cancellation_MR} that the total TTL coupling fully cancels if the centre of rotation lies on the beam's propagation axis, and the Gaussian beams have identical beam parameters. 
This holds also if transmissive components exist along the beam path (Sec.~\ref{sec:LPS_cancellation_tc}). 
In either case, the beam walk induced non-geometric TTL coupling cancels the geometric TTL effects.
However, if the centre of rotation is laterally shifted against the beam axis, the cancellation fails. Then, a significant geometric TTL coupling exists, that has no non-geometric counterpart.

We find an incomplete TTL cancellation in the case of unequal parameters. 
However, we can construct particular scenarios in which the cancellations holds again.
If the interfering beams have an identical waist position and a pivot in the centre of the waist, the geometric and the non-geometric signal cancel despite a possible arbitrary waist size mismatch, see Sec.~\ref{sec:LPS_cancellation_RS}.

With a dedicated lens system, one can image the centre of rotation of the beam onto its point of incidence at the detector \cite{SchusterPhD,Troebs2018}.
In this case, we find no geometric TTL coupling but non-geometric coupling terms due to the wavefront inequalities of the two beams. 
If imaging the point of rotation not exactly onto the detector surface or shifting the photodiode longitudinally, a small geometric coupling remains, which can for a suitable alignment counteract the non-geometric coupling \cite{Chwalla2020}. 
Therefore, such imaging systems can significantly suppress the observed TTL coupling and will be used for this purpose in the LISA mission (e.g.\ \cite{Chwalla2020}).

Another case of non-geometric TTL coupling without a geometric counterpart can be found if the points of incidence of the two beams do not coincide but have a static offset. 
We show in Sec.~\ref{sec: non_geom offset equal beam} that this offset breaks the wavefront symmetry even for equal beam parameters and generates linear non-geometric TTL coupling.
Thus, an intentional offset of one of the beams can counteract other linear TTL coupling effects without simultaneously changing the geometric TTL coupling.

Analogously, 
lateral jitter of the receiving system effectively changes the offset between the two beams at the detector without changing their geometric path length.
The resulting total lateral jitter coupling originates from non-geometric effects only, and is linear (Sec.~\ref{sec: non_geom offset equal beam RS lateral}).

This is different for the lateral jitter of a mirror, that moves into or out of the beam path.
This shortens or elongates the beams' optical path length.
On the other hand, the lateral jitter induces only negligible non-geometric TTL coupling effects since the reflected beam neither tilts nor significantly shifts in a lateral direction (Sec.~\ref{sec: non_geom offset equal beam TM lateral}).

For both considered systems, a lateral offset between the centre of rotation and the point of reflection (angular mirror jitter) or point of detection (angular receiver jitter), respectively, induces linear TTL coupling \cite{G21}.
This coupling is fully geometric since any non-geometric signal contributions are negligible, see Secs.~\ref{sec: SR-arbitrary_pivot_identical_Gaussians} and \ref{sec: non_geom offset equal beam TM}. 
Therefore, applying a lateral shift of the respective centre of rotation can be used to counteract other linear angular TTL coupling effects.

In summary, these presented TTL mechanisms can be used to counteract the overall TTL coupling even if the single underlying effects are unknown. 
This has been proven efficient in experiments (e.g.\ \cite{Troebs2018, Chwalla2020, LeePhD}).

\section{Conclusion}
\label{sec:conclusion}
Throughout this work, we have described TTL coupling as the
angular and translational motion of a mirror reflecting one of the interfered beams, or the jitter of a receiving system with respect to a received beam, coupling into the phase readout.
This coupling adds unwanted noise to the phase signal. 
The TTL coupling noise is an important noise source in precision interferometers, particularly in space interferometers, such as future space-based gravitational wave observatories like LISA, or the geodesy mission GRACE-FO and its successors.
In this work we categorised the different non-geometric TTL coupling mechanisms for the interference of two Gaussian beams.
We distinguish between the effects
originating from the characteristics of the wavefronts, and the detector geometry including different path length signal definitions. 

Wherever possible, we computed these non-geometric effects analytically from the overlap integral over the beams' electrical fields. 
The results agree with the numeric computations done by the simulation tool IfoCAD.

We summarised our key findings of the various non-geometric TTL coupling effects in Sec.~\ref{sec:summary} and Tab.~\ref{tab: TTL_NG_overview}.
Additionally, we combined these key findings with the geometric TTL results presented in \cite{G21} to estimate the total TTL coupling.
In Sec.~\ref{sec: overview}, we discussed in which cases the geometric and non-geometric TTL mechanisms cancel each other, or can intentionally be used to counteract other TTL effects for minimising the total TTL coupling.

Our findings can be a valuable tool for the suppression of TTL coupling noise either by design or realignment in an existing system.
Such a suppression is essential for a reduction of the TTL noise to a magnitude that can finally be removed by subtraction in post-processing \cite{armano2016sub,Giusteri2022}.

\ack
This work was made possible by funds of both the Deutsche Forschungsgemeinschaft (DFG) and the German Space Agency, DLR.
We gratefully acknowledge the Deutsche Forschungsgemeinschaft (DFG) for funding the Sonderforschungsbereich (SFB 1128: geo-Q) ``Relativistic Geodesy and Gravimetry with Quantum Sensors'', project A05 and all work contributions to this paper made by Sönke Schuster. 
Furthermore, we acknowledge DFG for funding the Clusters of Excellence PhoenixD (EXC 2122, Project ID 390833453) and QuantumFrontiers (EXC 2123, Project ID 390837967). 
Likewise, we gratefully acknowledge the German Space Agency, DLR and support by the Federal Ministry for Economic Affairs and Energy based on a resolution of the German Bundestag (FKZ 50OQ1801). 
Finally, we would like to acknowledge the Max Planck Society (MPG) for supporting the framework LEGACY on low-frequency gravitational wave astronomy, a cooperation between the Chinese Academy of Sciences (CAS) and the MPG (M.IF.A.QOP18098).

\bibliographystyle{unsrt}

\end{document}